\documentclass[traditabstract]{aa}

\usepackage{amsmath}
\usepackage{natbib}
\usepackage{graphicx}
\usepackage{txfonts}
\usepackage{xspace}
\usepackage{microtype}
\usepackage[colorlinks=true,linkcolor=red,urlcolor=black,citecolor=blue]{hyperref}

\bibpunct{(}{)}{;}{a}{}{,}

\graphicspath{{./figs/}}

\newcommand{\rxte}{\textsl{RXTE}\xspace}
\newcommand{\suzaku}{\textsl{Suzaku}\xspace}
\newcommand{\fu}{4U~0115+634\xspace}
\newcommand{\ergs}{\mathrm{erg}\,\mathrm{s}^{-1}\,\mathrm{cm}^{-2}}
\newcommand{\D}{\mathrm{d}}

\defcitealias{Mueller2013a}{M13}
\defcitealias{Iyer2015a}{I15}

\begin{document}

\title{The giant outburst of \fu in 2011 with \suzaku and \rxte}
\subtitle{Minimizing cyclotron line biases}

\author{
      Matthias Bissinger n{\'e} K\"uhnel \inst{1}
 \and Ingo Kreykenbohm \inst{1}
 \and Carlo Ferrigno \inst{2}
 \and Katja Pottschmidt \inst{3,4}
 \and Diana M. Marcu-Cheatham \inst{3,4}
 \and Felix F\"urst \inst{5}
 \and Richard E. Rothschild \inst{6}
 \and Peter Kretschmar \inst{5}
 \and Dmitry Klochkov \inst{7}
 \and Paul Hemphill \inst{8}
 \and Dominik Hertel \inst{1}
 \and Sebastian M\"uller \inst{1}
 \and Ekaterina Sokolova-Lapa \inst{1}
 \and Bosco Oruru \inst{9}
 \and Victoria Grinberg \inst{7}
 \and Silvia Mart\'inez-N\'u\~nez \inst{10}
 \and \mbox{Jos\'e M. Torrej\'on} \inst{11}
 \and Peter A. Becker \inst{12}
 \and Michael T. Wolff \inst{13}
 \and Ralf Ballhausen \inst{1}
 \and Fritz-Walter Schwarm \inst{1}
 \and J\"orn Wilms \inst{1}
}

\institute{
      Dr. Karl Remeis-Sternwarte \& ECAP, Universit\"at 
      Erlangen-N\"urnberg, Sternwartstr.~7, 96049 Bamberg, Germany
 \and ISDC Data Center for Astrophysics, University of Geneva, 16
      Chemin d'\'Ecogia, 1290 Versoix, Switzerland 
 \and Department of Physics, University of Maryland Baltimore County, Baltimore, MD 21250, USA 
 \and CRESST and NASA Goddard Space Flight Center, Astrophysics Science
Division, Code 661, Greenbelt, MD 20771, USA
 \and European Space Astronomy Centre
(ESA/ESAC), Science Operations Department, Villanueva de la Ca\~nada (Madrid), Spain 
 \and Center for Astrophysics and Space Sciences, University of 
      California, San Diego, La Jolla, CA 92093, USA
 \and Institut f\"ur Astronomie und Astrophysik, Universit\"at
      T\"ubingen, Sand 1, 72076 T\"ubingen, Germany 
 \and Massachusetts Institute of Technology, Kavli Institute for Astrophysics, Cambridge, MA 02139, USA
 \and Department of Physics, Makerere University, P. O. Box 7062, Kampala, Uganda
 \and Instituto de F\'isica de Cantabria, CSIC-Universidad de Cantabria,  39005 Santander, Spain
 \and Instituto Universitario de F\'isica Aplicada a las Ciencias y las 
      Tecnolog\'ias, University of Alicante, P.O.\ Box 99, 03080
      Alicante, Spain
 \and Department of Physics and Astronomy, George Mason University, Fairfax, VA
 \and Space Science Division, Naval Research Laboratory, Washington DC , 20375 USA
}

\abstract{We present an analysis of X-ray spectra of the high mass
  X-ray binary \fu as observed with \suzaku and \rxte in 2011 July,
  during the fading phase of a giant X-ray outburst. We used a
  continuum model consisting of an absorbed cutoff power-law and an
  ad-hoc Gaussian emission feature centered around 8.5\,keV, which we
  discuss to be due to cyclotron emission. Our
  results are consistent with a fundamental cyclotron absorption line
  centered at ${\sim}10.2$\,keV for all observed flux ranges. At the
  same time we rule out significant influence of the 8.5\,kev Gaussian
  on the CRSF parameters, which are not consistent with the cyclotron
  line energies and depths of previously reported flux-dependent
  descriptions. We also show that some continuum models can lead to
  artificial line-like residuals in the analyzed spectra, which are
  then misinterpreted as unphysically strong cyclotron
  lines. Specifically, our results do not support the existence of a
  previously claimed additional cyclotron feature at
  ${\sim}15$\,keV. Apart from these features, we find for the first
  time evidence for a He-like \ion{Fe}{XXV} emission line at
  ${\sim}6.7$\,keV and weak H-like \ion{Fe}{XXVI} emission close to
  ${\sim}7.0$\,keV.}

\date{Received 04 April 2019 / Accepted 13 Dezember 2019}

\let\Oldarcsec\arcsec
\renewcommand{\arcsec}{\Oldarcsec\xspace}

\maketitle \section{Introduction}

Be X-ray binaries (BeXRBs) consist of a neutron star and a Be-type
companion star and are a subclass of high mass X-ray binaries (HMXBs).
Mass transfer from the equatorial disk of the optical companion onto
the neutron star leads to violent X-ray outbursts. These outbursts are
classified either as regular, ``type~I'', outbursts associated with
the periastron passage of the neutron star, or giant, ``type~II
outbursts'' which are stochastic and less frequent \citep[see,
e.g.,][]{finger1997a} and due to strong mass transfer onto the compact
object. Once the transferred matter has reached the Alfv\'en radius of
the neutron star, it follows the magnetic field lines of the compact
object and is channeled onto the magnetic poles of the neutron star
\citep{Lamb1973a}. Due to the strong surface magnetic fields of
neutron stars on the order of a few $10^{12}$\,G, the energy of the
electrons perpendicular to the direction of the $B$-field is quantized
in discrete Landau levels. Resonant scattering processes of X-ray
photons with these electrons result in cyclotron resonance scattering
features (CRSFs, or cyclotron lines) in the X-ray spectra of some
X-ray pulsars. The cyclotron line energy is a function of the magnetic
field strength at the emission region. The respective energies of the
fundamental and harmonic CRSFs are given by the 12-$B$-12 rule, i.e.,
$E_\mathrm{CRSF,0} \sim 11.6\,\mathrm{keV}\times
B/10^{12}\,\mathrm{G}$
\citep[e.g.,][]{Meszaros1992a,Caballero2012a,Staubert2019a}, and
multiples. 

\fu consists of a neutron star with a pulse period of $\sim$3.6\,s
\citep{Cominsky1978a} which is in a $\sim$24.3\,d orbit
\citep{Bildsten1997a} with its Be-type companion V635~Cas
\citep{Negueruela2001a}. Since its discovery with the \textsl{Uhuru}
satellite \citep{Giacconi1972a}, it sporadically went into type~II
outbursts which were separated by several years of quiescence
\citep[][hereafter M13]{Mueller2013a}. These outbursts typically
lasted one to two months (e.g.,
\citealt{Tamura1992a,Heindl1999a,Nakajima2006a,Tsygankov2007a};
\citetalias{Mueller2013a}). The observed value of \fu's fundamental
cyclotron line energy of ${\sim}11$\,keV (\citealt{White1983a}; see
\citealt{Wheaton1979a} for the first detection of a CRSF in \fu,
which was actually the harmonic at ${\sim}$20.1\,keV) is rather low
compared to other known cyclotron line sources \citep[see,
  e.g.,][]{Staubert2019a}. In contrast, however, no other source than
\fu is known to show four additional cyclotron line harmonics in its
spectrum \citep{Heindl1999a,Santangelo1999a,Ferrigno2009a}, making
this source an ideal target for testing theories on the luminosity
dependency of the cyclotron line energy, which traces the magnetic
field at the location where most of the X-rays are emitted.

As widely discussed in the literature, there are at least two
accretion regimes for X-ray pulsars, which are separated by the
so-called critical luminosity $L_\text{crit}$ \citep[e.g.,][and
references
therein]{Basko1976a,Staubert2007a,Becker2012a,Mushtukov2015a,Postnov2015a}.
Sources at lower accretion rates, that is, sources with X-ray
luminosities $L_\text{X} \lesssim L_\text{crit}$ are expected to
exhibit a positive correlation between $E_0$ and $L_\text{X}$ (e.g.,
Her~X-1, \citealt{Staubert2007a}, GX~304$-$1, \citealt{Klochkov2012a},
\citealt{Rothschild2017a}). Here, as the mass accretion rate increases
the accretion column decreases in height, and therefore the X-rays are
emitted in regions where the local $B$-field is larger. In contrast,
for sources where $L_\text{X}\gtrsim L_\text{crit}$ a negative
correlation is expected \citep[e.g., V~0332+53,][]{Tsygankov2006a,
  Mowlavi2006a}, that is, higher mass accretion rates lead to larger
accretion columns. \fu had originally been seen as the poster child
for this group of high luminosity sources
\citep[e.g.,][]{Nakajima2006a, Tsygankov2007a, Mueller2010a, Li2012a}.
However, \citetalias{Mueller2013a} showed that the inferred behavior
of the cyclotron line in \fu strongly depends on the choice of the
broadband continuum model used to describe the X-ray spectra.
Specifically, \citetalias{Mueller2013a} showed that if the popular
negative and positive exponent power-law (NPEX) model is used to
describe data from \fu, strong cyclotron lines are found. The model
components corresponding to these lines, however, do not describe the
actual shapes of the cyclotron lines, but rather erroneously model
part of the underlying continuum emission. Since the continuum shape
of \fu is luminosity-dependent, the result is the strong cyclotron
line variability claimed in many earlier papers. In contrast, the
model used by \citetalias{Mueller2013a} results in a constant
cyclotron line energy for all observed flux levels in \fu, i.e., the
previously observed anticorrelation vanishes. The dependency of the
cyclotron line parameters on the continuum model was later confirmed
by \citet{boldin2013a}.

In this paper we present an analysis of data taken during a giant
outburst of \fu in 2011 with \suzaku and \rxte. These data were
previously analyzed by \citet{Iyer2015a}, who claim the detection of
complex and strong cyclotron lines. In Sect.~\ref{sec:obs} we
summarize the observations and describe the data extraction process.
In Sect.~\ref{sec:specana} we present the results of the spectral
analysis of the X-ray spectra. Specifically, we show in
Sect.~\ref{sec:discuss} that unphysically strong and complex cyclotron
lines are not needed to describe the data, and that these strong lines
imply a much larger intrinsic source continuum flux than
observed. Using the continuum model of \citetalias{Mueller2013a}, on
the other hand, results in more physical cyclotron lines that are
consistent with earlier measurements. We summarize our results and
draw conclusions in Sect.~\ref{sec:summary}.

\begin{figure}
 \includegraphics[width=\columnwidth]{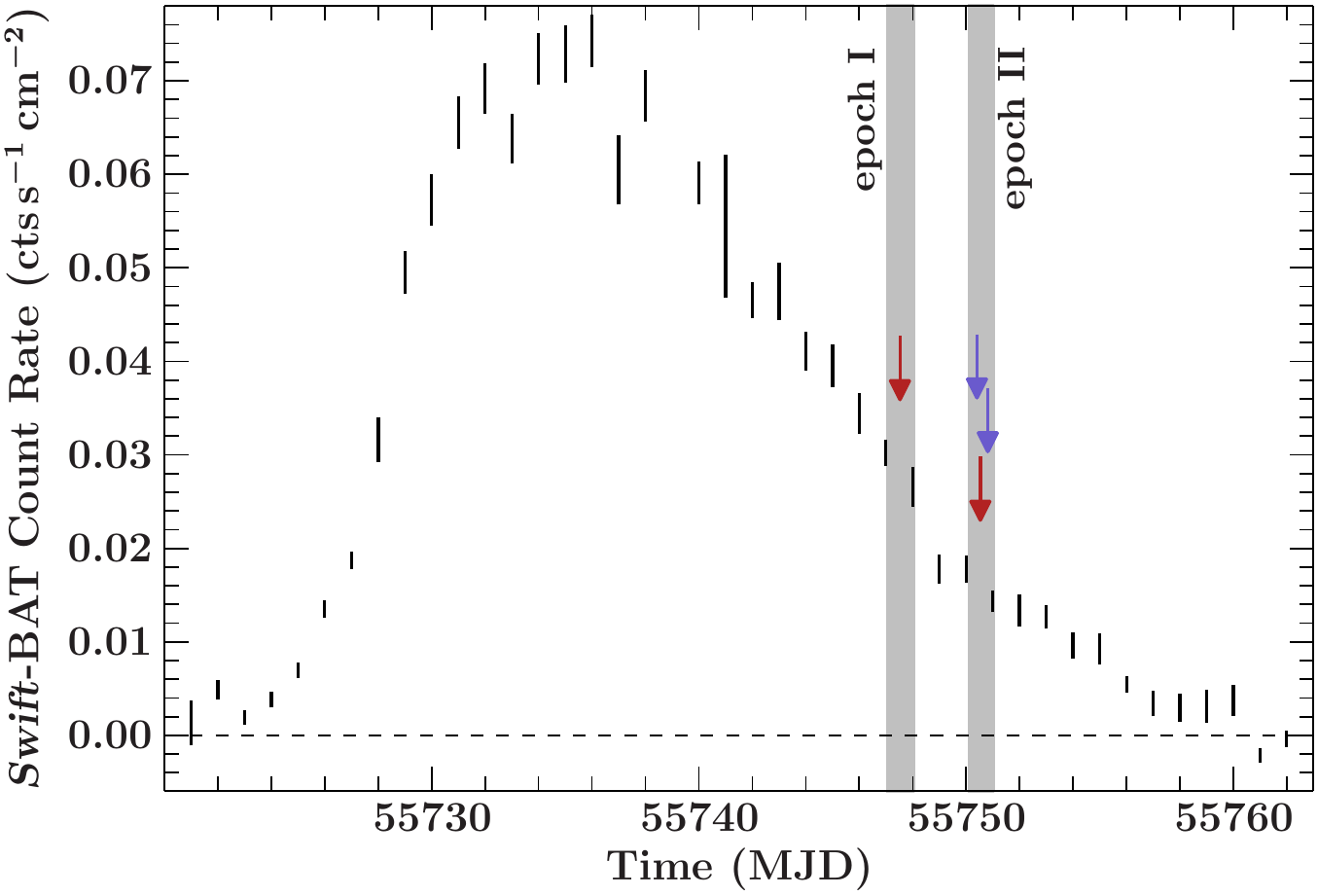}
 \caption{\textsl{Swift}-BAT daily lightcurve of \fu (15--50\,keV)
   during its 2011 June/July outburst. The arrows mark the midtimes of
   the observations by \suzaku (red) and \rxte (blue). The defined
   data epochs~I and II are indicated by vertical gray bands.}
 \label{fig:lc}
\end{figure}

\section{Observations and data reduction}\label{sec:obs}

The 2011 outburst of \fu started on June 10 (MJD 55722), after more
than three years of quiescence, when MAXI-GSC detected a significant
increase of the X-ray flux \citep{Yamamoto2011b}. The outburst lasted
about 40\,days and exceeded a flux of $\sim$300\,mCrab in the
15--50\,keV band of the BAT instrument onboard the \textsl{Neil
  Gehrels Swift Observatory} \citep[Fig.~\ref{fig:lc};][]{Krimm2013a}.
The source was observed by \suzaku on 2011 July 5 (Obs-ID 406048010,
24\,ksec, epoch~I in the remainder of this work) and 2011 July 8
(Obs-ID 406049010, 42\,ksec, epoch~II in the remainder of this work).
During the latter observation \rxte observed \fu twice (see
Fig.~\ref{fig:lc}; Obs-IDs 96032-01-04-00 and -01, 13\,ksec and
1\,ksec, respectively).

\suzaku had two main detector assemblies. The X-Ray Imaging
Spectrometer \citep[XIS,][]{Koyama2007a} consisted of four Si-based
X-ray charge coupled device (CCD) cameras, XIS0--XIS3, where XIS2 was
no longer operational since a micro-meteorite hit in
2006\footnote{\url{https://heasarc.gsfc.nasa.gov/docs/suzaku/news/xis2.html}}.
The three remaining CCDs covered the energy range between 0.2 and 12
keV. The Hard X-Ray Detector \citep[HXD,][]{Takahashi2007a} extended
the energy range of \suzaku up to $\sim$600\,keV and consisted of two
instruments: the PIN silicon diodes (PIN) were mainly sensitive below
$\sim$60\,keV, while the GSO/BGO phoswich counter (GSO) was sensitive
above $\sim$40\,keV.

The event data of both observations were reprocessed by
\texttt{aepipeline} as provided by the \textsc{HEASOFT} software
package (v.~6.25). The version of the HEASARC calibration database
(CALDB) used for the X-ray telescopes (XRT) was 20110630, for the XIS
20181010, and for the HXD 20110913.

First, we corrected the attitude using \texttt{aeattcor2} and applied
the result to the XIS events using \texttt{xiscoord}. The images
extracted using \texttt{xselect} were checked for pile-up using
\texttt{pileest} called with a grade migration parameter (alpha) of
0.5. We used the pile-up-images to create extraction regions for each
XIS and editing mode as follows: the source region was represented by
an annulus with a 90\,\arcsec outer radius. To avoid pile-up fractions
above 4\% the inner radius had to be 45\arcsec for observation
406048010 and 55\arcsec for 406049010. The larger excluding radius is
a result of the overall increased pile-up fraction during the second
observation, although the source was weaker compared to the first
observation (compare Fig.~\ref{fig:lc}). This is due to the fact that
the XIS normal clock mode options were set to 1/4 window and 1\,s
burst during the first observation, while the burst option was removed
during the second, which doubled the exposure time per frame from 1\,s
to 2\,s. The two background regions for each XIS and mode were circles
with a radius of 60\arcsec located near the borders of the chips.
After having extracted the events filtered by these regions using
\texttt{xselect} we added the spectra of the different editing modes.

For spectral analysis, we used data from XIS0, XIS1, and XIS3 in the
energy range 0.8--9\,keV. The recent calibration update (20181010)
focused on improving known calibration features around the Si K edge
at ${\sim}$1.84\,keV \citep{Okazaki2018a}. This update introduced a
jump in the relation of the incident photon energy and pulse height,
which was attributed to charge losses in the depletion layer of the
CCDs. We investigated the updated calibration and found that
significant residuals around the Si K edge are no longer present in
the spectra of XIS0 and XIS3 for both considered observations (see
Fig.~\ref{fig:XIScalib}). Furthermore, the new calibration seems to
reduce calibration features around the Au M edge at 2.22\,keV in these
XISs as well\footnote{A similar improvement can be seen in the
  document describing the latest XIS CALDB files, see
  \url{https://heasarc.gsfc.nasa.gov/docs/heasarc/caldb/suzaku/docs/xis/20180807_sikedge_xisrmfgen.pdf}.}. Unfortunately,
the residuals around both features are still present in XIS1 (see
Fig.~\ref{fig:XIScalib}). Thus, we excluded the energy ranges
1.73--1.95\,keV and 2.16--2.37\,keV for XIS1 during the analysis. The
spectra were rebinned according to \citet{Nowak2011a}, such that each
energy bin has a signal-to-noise ratio (SNR) of 8 at least and a
minimum number of channels close to the half-width half-maximum of the
spectral resolution. Data from the PIN instrument were used in the
energy range 16--55\,keV for spectral fitting and rebinned to a SNR of
5.

The non X-ray background (NXB) spectral extraction based on model
events and the cosmic X-ray background (CXB) simulation were performed
using the \texttt{hxdpinxbpi} tool. To get a background spectrum of the
PIN, we combined the version 2.0 of the ``tuned'' NXB and the CXB,
which was simulated using the model as described in
\citet{Boldt1987a}. As PIN response we used the epoch 11 response file
(20110601) for the XIS nominal position. Because of an insufficient
SNR, we did not use the data from the GSO instrument of the HXD.

For the analysis of \rxte data, data from the top layer of unit 2 of
the Proportional Counter Array \citep[PCA,][]{Jahoda2006a} were
used. Since the launch of \rxte in 1995 until the 2011 outburst of \fu
the calibration of the remaining Proportional Counter Units (PCU)
became complicated with the aging of the instruments. PCU2 is,
however, still known as the best calibrated one \citep{Jahoda2006a}.
The spectra of the both observations, which were performed during the
second \suzaku observation (epoch~II), were reduced using the
\textsc{HEASOFT} software package (v. 6.25) and standard data
reduction pipelines \citep[and references therein]{wilms2006a}. The
event times were filtered excluding the first 30 minutes since the
start of the South Atlantic Anomaly (SAA) passages and with an
elevation angle of greater than $10^\circ$ above the Earth's
limb. After having combined the spectra of both observations, we added
1\% systematic uncertainties to all channels below 7\,keV (see
Fig.~\ref{fig:PCAcalib}) due to calibration issues near the Xe L-edge
at 4.5\,keV, uncertainties in the PCA background estimation
\citep{Jahoda2006a}, or possible inaccurately assigned energies to the
channel boundaries \citep{Garcia2014a}. Channels in the energy range
of 3.5 keV--50\,keV were used for spectral analysis and were grouped
into bins with a minimum SNR of 1. Data from the High Energy X-ray
Timing Experiment \citep[HEXTE,][]{rothschild1998a} were not used here
since the rocking mechanism, which is crucial to measure the detector
background, was switched off completely in 2010 April.

\section{Spectral analysis}\label{sec:specana}
Data modeling in this work was performed with the \textit{Interactive
  Spectral Interpretation System} \citep[ISIS,][]{Houck2000a} version
1.6.2-41. All uncertainties are given at the 90\% confidence level unless
 stated otherwise.

\begin{table*}
  \centering
  \caption{Final parameters of the spectral analysis. Both cyclotron
    line models \texttt{CYCLABS} and \texttt{GABS} have been fitted
    independently on top of the same continuum and iron line
    models. See Sect.~\ref{sec:models} for the definition of
    parameters and applied models. The second table below lists the
    parameters using a single broad iron line instead of three narrow
    ones.}
  \label{tab:fitpars} 
  \renewcommand{\arraystretch}{1.3}
  \begin{tabular}{ll|ll|ll}
    \hline\hline
               &      & \multicolumn{2}{c|}{\texttt{CYCLABS}} & \multicolumn{2}{c}{\texttt{GABS}} \\
     Parameter & Unit & Epoch~I & Epoch~II & Epoch~I & Epoch~II \\
    \hline
    \csname @@input\endcsname paramtable \hline
    \multicolumn{6}{l}{}\\
    \multicolumn{6}{l}{broad iron line}\\
    \hline
    $\mathrm{EW}_\mathrm{Fe}$
& eV
& $55_{-14}^{+17}$
& $22_{-6}^{+7}$
& $52_{-14}^{+17}$
& $16_{-5}^{+6}$
\\
$E_\mathrm{Fe}$
& keV
& $6.61_{-0.07}^{+0.07}$
& $6.56_{-0.06}^{+0.06}$
& $6.61_{-0.07}^{+0.07}$
& $6.55_{-0.06}^{+0.07}$
\\
$\sigma_\mathrm{Fe}$
& eV
& $410_{-90}^{+100}$
& $210_{-70}^{+90}$
& $400_{-90}^{+100}$
& $170_{-60}^{+80}$
\\

    \hline
  \end{tabular}
  \tablefoot{
     Uncertainties and upper limits are given at the 90\% confidence level.\\
     \tablefoottext{$\dagger$}{Fixed.}\\
     \tablefoottext{a}{Unabsorbed \texttt{CutoffPL} flux in the 3--50\,keV
                       energy band.}\\
     \tablefoottext{b}{Unabsorbed, bolometric flux in the 8.5\,keV Gaussian.}\\
     \tablefoottext{c}{Fixed to the value of epoch~II.}\\
  }
\end{table*}

\subsection{Spectral model}\label{sec:models}

We used the same approach as \citetalias{Mueller2013a} to model the
broadband X-ray spectra of \fu: a cutoff power-law, called
\texttt{CutoffPL} in ISIS and XSPEC \citep{Arnaud1996a}, which has the
form
\begin{equation}\label{eq:cutoffpl}
 \texttt{CutoffPL}(E)\propto E^{-\Gamma}\exp(-E/E_\text{fold}),
\end{equation}
with the photon index, $\Gamma$, and the folding energy
$E_\text{fold}$. We normalized the \texttt{CutoffPL} model to the
unabsorbed photon flux, $F_\mathrm{PL}$, integrated over the
3--50\,keV energy range. The cutoff power-law was modified by a broad
Gaussian emission feature around 8.5\,keV. This is more commonly known
as the ``10\,keV feature'' and has been observed in \fu before
\citep[see, e.g.,][]{Ferrigno2009a,Mueller2012a} and in many other
X-ray pulsars \citep[see, e.g.,][]{Coburn2002a}, even though the
origin of this feature is still under discussion and remains unclear.
Here, the normalization of the Gaussian corresponds to its bolometric
photon flux, $F_\mathrm{10\,keV}$ (technically derived by integrating
over the 1--25\,keV energy range).

We modeled the various cyclotron lines present in the spectra of \fu
by two models implementing different line profiles of the absorption
line-like features. The first model, \texttt{CYCLABS}, represents
pseudo-Lorentzian profiles given by \citep{Mihara1990a}
\begin{equation}\label{eq:crsf-cyclabs}
  \texttt{CYCLABS}(E) = \exp\left(-\frac{\tau_\text{CRSF}(W_\text{CRSF}E/E_\text{CRSF})^2}
    {(E-E_\text{CRSF})^2+W_\text{CRSF}^2}\right),
\end{equation}
where $E_\text{CRSF}$ is the centroid energy, $W_\text{CRSF}$ is the
width of the feature, and $\tau_\text{CRSF}$ is the optical depth. The
second model, \texttt{GABS}, implements a Gaussian optical depth
profile given
by\footnote{\url{https://heasarc.gsfc.nasa.gov/xanadu/xspec/manual/node240.html}}
\begin{equation}\label{eq:crsf-gabs}
  \texttt{GABS}(E) = \exp\left(-\frac{D_\text{CRSF}}{\sqrt{2 \pi} \sigma_\text{CRSF}} \exp\left( -\frac{(E-E_\text{CRSF})^2}{2 \sigma_\text{CRSF}^2} \right) \right),
\end{equation}
where $D_\text{CRSF}$ is the strength and $\sigma_\text{CRSF}$ the
Gaussian width of the cyclotron line. In the remainder of this paper,
we label the parameters of the cyclotron lines for both models with
the number of the respective harmonic, where 0 denotes the fundamental
line.

We emphasize that the cyclotron line strength, $D_\text{CRSF}$, of the
\texttt{GABS} model is \emph{not} the optical depth at line center. At
$E = E_\text{CRSF}$ the optical depth is
\begin{equation}\label{eq:cyclabs-gabs}
  \tau_\text{CRSF} = \frac{D_\text{CRSF}}{\sqrt{2 \pi} \sigma_\text{CRSF}}
\end{equation}
which is equivalent to the optical depth parameter of the
\texttt{CYCLABS} model, $\tau_\text{CRSF}$. In a similar manner, the
width parameters $W_\text{CRSF}$ and $\sigma_\text{CRSF}$ are related
to each other, but not identical.

The \suzaku-XIS spectra also show source intrinsic emission lines of
iron between 6\,keV and 7\,keV. We modeled these lines using Gaussians
with their centroid energies, $E_\mathrm{Fe}$, widths,
$\sigma_\mathrm{Fe}$, and equivalent widths, $\mathrm{EW}_\mathrm{Fe}$
as a measure of their X-ray flux. We label these parameters further
according to the observed transition or ion, e.g., K$\alpha$ or
XXV. The calculation of the equivalent widths was based on the
unabsorbed model flux, i.e., the source intrinsic flux.

In order to account for the interstellar absorption affecting
primarily the XIS spectra, we used \texttt{tbnew}, an updated version
of \texttt{tbabs}\footnote{see
  http://pulsar.sternwarte.uni-erlangen.de/wilms/research/tbabs/},
incorporating abundances by \citet{Wilms2000a} and cross sections by
\citet{Verner1995a}. The equivalent hydrogen column density is denoted
as $N_\text{H}$.

Since the flux normalizations of the different instruments are not
perfectly known, we introduced the cross-calibration constants
$c_\text{XIS0}$, $c_\text{XIS1}$, $c_\text{XIS3}$ and
$c_\text{PIN}$. We chose XIS3 to be the reference instrument and thus
set $c_\text{XIS3}=1$. To account for the cross calibration between
the \suzaku and \rxte data, we introduced the constant
$c_\text{PCA}$ referring to the combined spectrum of the two PCA
observations.

\subsection{Modeling strategy}\label{sec:modeling}

\begin{figure*}
  \begin{minipage}{\columnwidth}
    \includegraphics[width=\columnwidth]{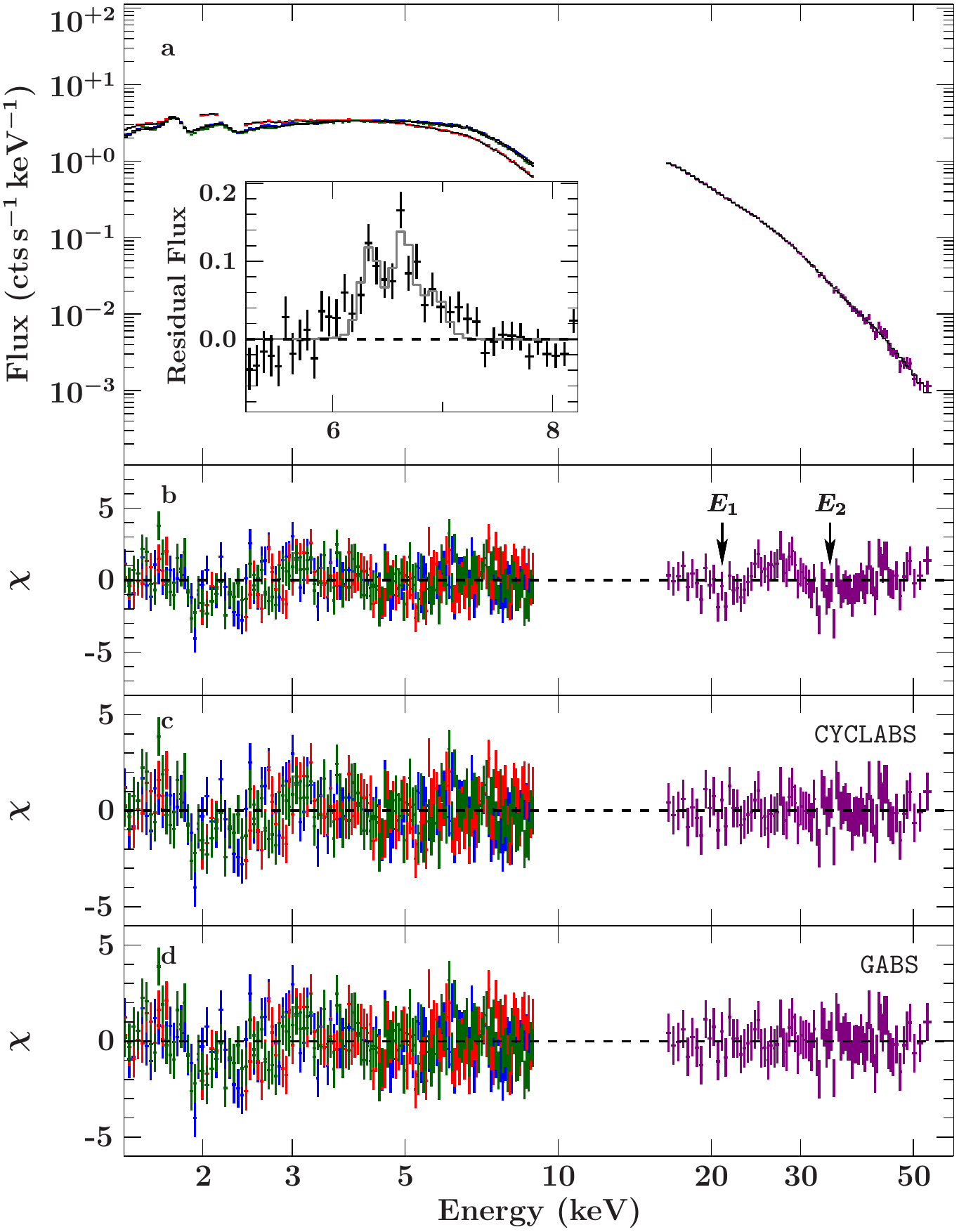}
  \end{minipage}\hfill\begin{minipage}{\columnwidth}
    \includegraphics[width=\columnwidth]{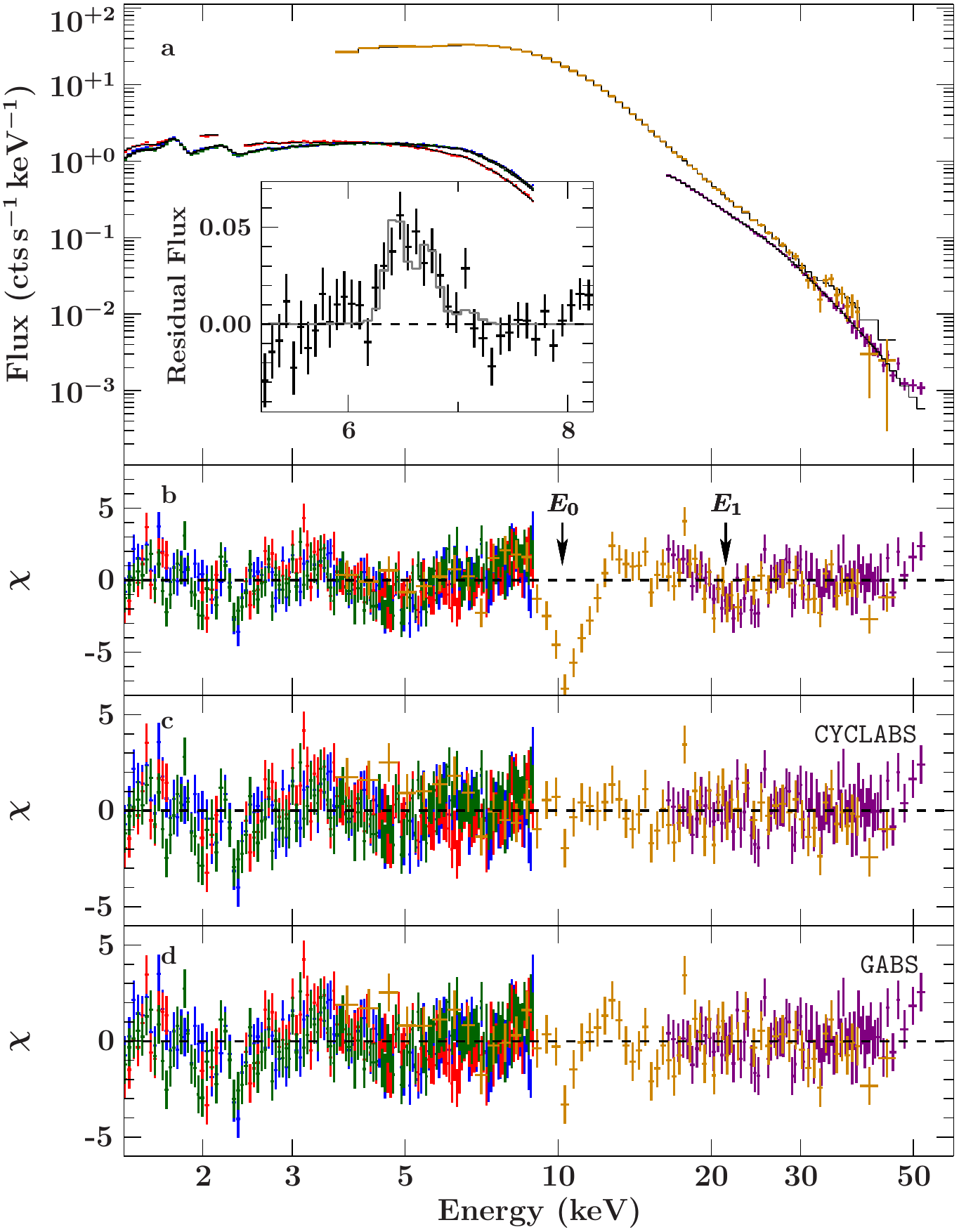}
  \end{minipage}
  \caption{Observed spectra and best-fit model of epoch~I (left) and
    epoch~II (right). Panels \textbf{a} show the \suzaku spectra
    (XIS0: green; XIS1: red; XIS3: blue; PIN: purple) and in case of
    epoch~II (right) the combined \rxte spectrum (PCA: orange). The
    black line is the best-fit model. Each \textbf{inset} zooms into
    the iron line region between 5 and 8\,keV and shows the residual
    flux with respect to a model without any fluorescence lines. The
    modeled fluorescence lines are shown in gray. Panel \textbf{b}
    shows the residuals of a fit to a model without the cyclotron
    lines included. The arrows labeled $E_\mathrm{n}$ mark the
    best-fit position of the $n^\mathrm{th}$ cyclotron line (if
    significant within the error bars, see Table~\ref{tab:fitpars}).
    Panels \textbf{c} and \textbf{d} show the residuals of our
    best-fits using \texttt{CYCLABS} and \texttt{GABS} to model the
    cyclotron lines, respectively.}
  \label{fig:spec}
\end{figure*}

The fundamental cyclotron line of \fu is known to be around 11\,keV
\citep{White1983a}. Unfortunately, this energy is within the data gap
between \suzaku-XIS and -PIN, i.e., between 9 and 16\,keV. Hence, in
order to avoid a fit degeneracy between this cyclotron line and the
remaining spectral parameters of \fu, we first performed a
simultaneous fit of epoch~II, where \rxte observed the source twice
during the corresponding \suzaku observation.

We used the continuum parameters of \citetalias{Mueller2013a} as
initial values for the fit. Using a pure continuum model without any
emission lines or CRSFs, emission line-like residuals at 6--7\,keV and
broad absorption line-like residuals around $\sim$11\,keV,
$\sim$22\,keV and $\sim$33\,keV are visible (see insets and panels c
of Fig.~\ref{fig:spec}, respectively). The absorption features
represent the fundamental CRSF and two of its higher harmonics. Thus,
we added three cyclotron line features to our model using two
different descriptions, the \texttt{CYCLABS}-model
(Eq.~\ref{eq:crsf-cyclabs}) and the \texttt{GABS}-model
(Eq.~\ref{eq:crsf-gabs}), resulting in two separate fits. Following
\citepalias{Mueller2013a} we fixed the width in both fits
($W_\mathrm{CRSF}$ and $\sigma_\mathrm{CRSF}$, respectively) of the
fundamental, the first, and second harmonic CRSF to 2\,keV, 2\,keV and
4\,keV, respectively, to avoid parameter degeneracies with the
continuum parameters.

In order to model the emission of iron at 6--7\,keV we added a single
Gaussian to the model. After an initial fit we found a centroid energy
of $E_\mathrm{Fe} = 6.56^{+0.06}_{-0.06}$\,keV, which is not
consistent with emission from neutral iron at 6.4\,keV. Its width of
$\sigma_\mathrm{Fe} = 210^{+90}_{-70}$\,eV is too broad to be
intrinsically narrow and, thus, is due to the astrophysics of the
source. In epoch~I the Gaussian's width of $\sigma_\mathrm{Fe} =
410^{+100}_{-90}$\,eV is even broader. Such broad iron lines are
usually not observed in HMXBs ($\sigma_\mathrm{Fe} \lesssim 150$\,eV,
see, e.g., \citealt{gimenez-garcia2015a} or \citealt{Torrejon2010a}
for comprehensive studies). In a \textsl{Chandra} observation of the
BeXRB A~0535+262, \citet{Reynolds2010a} discovered fluorescence lines
from highly ionized iron besides a narrow emission line from neutral
iron. In a spectrum recorded by an X-ray instrument with lower
spectral resolution, these lines would blend into a single broad
emission line at a higher centroid energy than 6.4\,keV as we found
here for \fu. We thus replaced the single broad Gaussian with two
Gaussians with a fixed width of 1\,eV, i.e., the emission lines can be
considered to be intrinsically narrow as expected. The centroid
energies of these lines of ${\sim}6.4$\,keV and ${\sim}6.7$\,keV (see
Table~\ref{tab:fitpars}) are consistent with fluorescent K$\alpha$
emission from neutral Fe and He-like \ion{Fe}{xxv},
respectively. Finally, we accounted for the contribution of an iron
K$\beta$ transition by adding a Gaussian with its centroid energy and
equivalent width tied relative to the fit-parameters of the neutral
K$\alpha$ emission line ($E_\mathrm{FeK\,\beta} =
E_\mathrm{Fe\,K\alpha} + 0.656$\,keV and
$\mathrm{EW}_\mathrm{Fe\,K\beta} = 0.16 \times
\mathrm{EW}_\mathrm{Fe\,K\alpha}$, respectively).

Our final model is able to describe the combined fit of the \suzaku-
and \rxte-data of epoch~II well ($\chi^2_\mathrm{red} = 1.44$ with 574
degrees of freedom, dof, using \texttt{CYCLABS}; $\chi^2_\mathrm{red}
= 1.51$ with 574 dof using \texttt{GABS}). The best-fit parameters
are listed in Table~\ref{tab:fitpars}.

We then performed a fit to the epoch~I data, consisting of a single
\suzaku observation, with the initial parameters provided by the
best-fit model from epoch~II. We fixed the parameters of the
fundamental CRSF, as it lies in the gap between the XIS and PIN. While
the epoch~II model fits the epoch~I data acceptably well for the most
part, there are some additional weak line-like residuals around 7\,keV
in the XIS spectrum, requiring a fourth narrow Gaussian to be added to
the model. We associate this emission with H-like \ion{Fe}{xxvi}. The
final best-fit to the data of epoch~I again results in a good
description of the data ($\chi^2_\mathrm{red} = 1.29$ with 516 dof
using \texttt{CYCLABS}; $\chi^2_\mathrm{red} = 1.29$ with 516 dof
using \texttt{GABS})).

See Table~\ref{tab:fitpars} for the final fit parameters for both
epochs and Fig.~\ref{fig:spec} for the observed spectra and
corresponding model.

\subsection{Significances \& parameter degeneracies}

In order to estimate the significance of the emission feature of
ionized iron, we simulated $N=4000$ spectra for each epoch on the
basis of our model while excluding the corresponding Gaussian
component. The neutral $K\alpha$ line has been included in the
simulated spectra and for the significance calculation of
\ion{Fe}{XXVI} in epoch~1 the \ion{Fe}{XXV} line has been included as
well. Each simulated spectrum was then fitted once without and once
including the Gaussian component. An improvement in the fit goodness,
$\Delta \chi^2_\mathrm{sim}$, higher than the observed $\Delta \chi^2$
was then counted as a false positive detection. No false positives
were found for He-like \ion{Fe}{xxv} around 6.7\,keV in epoch~I and
epoch~II ($\Delta \chi^2 = 65.86$ with 516 dof and $\Delta \chi^2 =
24.51$ with 574 dof, respectively) corresponding to significances of
$\ge 3.66\sigma$ for both epochs. For H-like \ion{Fe}{xxvi} in
epoch~I, 16 false positives were found in the simulation ($\Delta
\chi^2 = 8.9$ with 516 dof), which corresponds to a significance of
${\sim}2.87\sigma$.

\begin{figure}
 \includegraphics[width=\columnwidth]{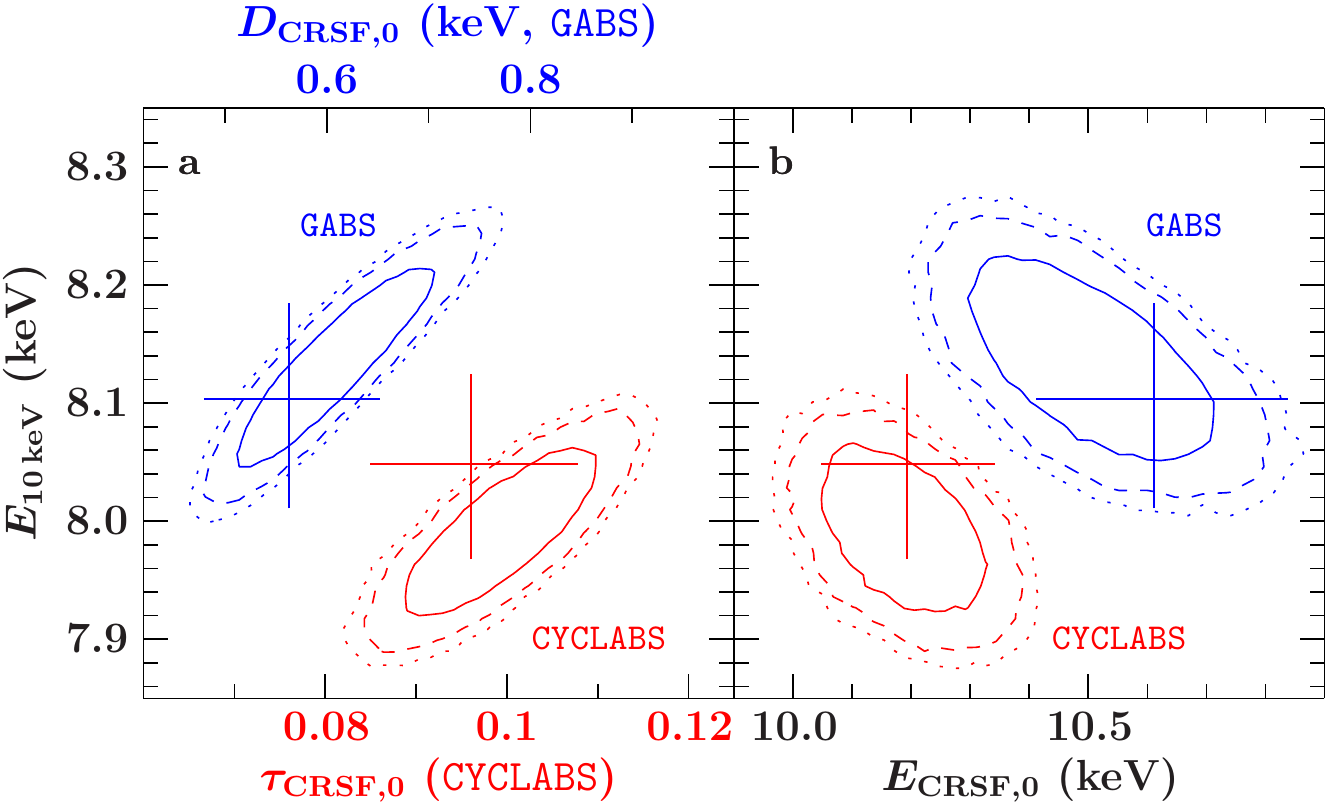}
 \caption{2D-probability contours between the fitted parameters of the
   fundamental cyclotron line (\textbf{a} depth,
   $\tau_\mathrm{CRSF,0}$ (\texttt{CYCLABS}) and $D_\mathrm{CRSF,0}$
   (\texttt{GABS}); \textbf{b} energy, $E_\mathrm{CRSF,0}$) and the
   centroid energy, $E_\mathrm{10\,keV}$, of the 10\,keV feature
   calculated by a MCMC method for epoch~II and both CRSF models
   (\texttt{CYCLABS}, red; \texttt{GABS}, blue). The contours
   correspond to the 68\% (solid line), 90\% (dashed line), and 99\%
   confidence level (dotted line) according to the MCMC walker
   distribution. The crosses are our best-fit parameters and their
   uncertainties using $\chi^2$-minimization (see
   Table~\ref{tab:fitpars}).}
 \label{fig:MCMCmap}
\end{figure}

We investigated the strength of possible parameter degeneracies
between the fundamental cyclotron line around 10.2\,keV and the
8.5\,keV Gaussian, i.e., the 10\,keV feature (see
Sect.~\ref{sec:discuss:10keV} for its detailed discussion). This
investigation is possible for epoch~II only due to the fixed CRSF
parameters during epoch~I. To reveal any possible degeneracies we have
sampled the parameter space using a Markov chain Monte Carlo (MCMC)
method after \citet{goodman2010a}. We used the implementation
\texttt{emcee} by \citet{foreman-mackey2013a}, which has been ported
into ISIS by M.~A.~Nowak\footnote{The ISIS implementation of
  \texttt{emcee} is distributed via the ISISscripts at
  \url{http://www.sternwarte.uni-erlangen.de/isis}}. We used 2\,000
iteration steps and 100 walkers per free parameter, which were
initially distributed randomly within the allowed parameter space. The
algorithm converged after 1200 steps and, thus, the first 60\% of the
parameter chain were ignored in the following. After having
investigated all possible 2D-probability distributions of the chain,
we do not find a significant degeneracy between the Gaussian's width
or flux with the CRSFs parameters. The only significant degeneracies
we find as show in Fig.~\ref{fig:MCMCmap} were present between the
Gaussian's centroid energy, $E_\mathrm{10\,keV}$, and the fundamental
cyclotron line energy, $E_\mathrm{CRSF,0}$, and its depth,
$\tau_\mathrm{CRSF,0}$ (\texttt{CYCLABS}) or $D_\mathrm{CRSF,0}$
(\texttt{GABS}). The best-fit parameter values found by
$\chi^2$-minimization (crosses in Fig.~\ref{fig:MCMCmap}) are in very
good agreement with the MCMC contours. Their respective uncertainties
as derived from the $\chi^2$-landscape are, however, too symmetric
compared to the elliptical contour shapes. For the remainder of this
work, we therefore consider the uncertainties of the Gaussian's
centroid energy, $E_\mathrm{10\,keV}$, the fundamental cyclotron line
energy, $E_\mathrm{CRSF,0}$, and its depth, $\tau_\mathrm{CRSF,0}$,
during epoch~II to be larger by a factor of 2 in order to cover the
asymmetric MCMC contours. We note, however, that even without
increasing the uncertainties the degeneracies do not bias our
discussions and conclusions in the following sections. The contours
further show a shift of $+0.15$\,keV in the Gaussian's centroid energy
when using the \texttt{GABS} CRSF model instead of the
\texttt{CYCLABS} model. This is, however, negligible compared to the
uncertainty of the Gaussian's centroid energy and does not bias our
discussions either.

\section{Discussion}
\label{sec:discuss}

During the following discussions the luminosity, $L$, and the
mass-accretion rate, $\dot{M}$, of \fu are needed. From the measured
fluxes $F_\mathrm{PL}$ and $F_\mathrm{10\,keV}$ of the power-law and
the 10\,keV feature, respectively (see Table~\ref{tab:fitpars}), we
calculated the total \mbox{3--50\,keV} luminosity of
$3.5{\times}10^{37}\,\mathrm{erg}\,\mathrm{s}^{-1}$ and
$2.5{\times}10^{37}\,\mathrm{erg}\,\mathrm{s}^{-1}$ during epoch~I and
epoch~II, respectively, using the distance of $7.2^{+1.5}_{-1.1}$\,kpc
determined by \cite{bailerjones2018a} from the Gaia DR2 parallax
measurements and consistent with earlier distance estimates of 7\,kpc
\citep{Negueruela2001a,Riquelme2012a}. Assuming that all potential
energy is released into radiation, we calculated the mass-accretion
rate,
\begin{equation}
  \dot{M}=\frac{L R}{G M}\quad{,}
\end{equation}
during epoch~I and epoch~II to be $1.9 \times
10^{17}\,\mathrm{g}\,\mathrm{s}^{-1}$ and $1.3 \times
10^{17}\,\mathrm{g}\,\mathrm{s}^{-1}$, respectively, assuming a
neutron star radius of $R=10^6\,\mathrm{cm}$ and a mass of
$M=1.4\,\mathrm{M}_\odot$ of the neutron star.
 
\subsection{Iron line region}

The feature causing the emission residuals at $\sim$6--7\,keV,
displayed in more detail in the inset of Fig.~\ref{fig:spec}, has
previously been modeled by a single Gaussian using data of coarse
spectral resolution (\rxte-PCA, $E/\Delta E \approx 6$ in the iron
line region). The feature has been interpreted as a narrow iron
K$\alpha$ fluorescence line at $\sim$6.4\,keV with an equivalent width
of $\sim$50--60\,eV \citepalias{Mueller2013a} or larger
\citep[e.g.,][]{Tsygankov2007a, Li2012a, boldin2013a}.

Due to the improved energy resolution of the \suzaku-XIS spectra
($E/\Delta E \approx 50$ in the iron line region) compared to earlier
observations, we found the line centroid energy blue-shifted by
$160\text{--}210$\,eV with respect to neutral iron K$\alpha$ emission
and an intrinsically broadening of the line by
$\sigma_\mathrm{Fe}{\approx}200\text{--}400\,\mathrm{eV}$ (see
Table~\ref{tab:fitpars} for the values of both data epochs). As argued
during the spectral analysis (Sect.~\ref{sec:modeling}) we proposed
that this emission actually is a blend of 2--3 emission lines by
highly ionized iron rather than a single blue-shifted and broad
neutral iron K$\alpha$ line. In an alternative scenario the neutral
iron line would originate from material around the neutron star, e.g.,
an accretion disk where its intrinsical rotation causes a broadening
of the line. In the following we will show that this scenario is,
however, in contradiction to the observed blue-shift of the line
centroid energy.

Emission lines emerging from an accretion disk should be Doppler
shifted accordingly to the binary motion. The orbital
velocity of the neutron star projected on the line of sight,
\begin{equation}\label{eq:vrad}
  v(t) = \frac{2 \pi a \sin i}{P_\mathrm{orb} \sqrt{1-e^2}}
  (\cos(\theta(t) + \omega) + e \cos \omega)\quad{,}
\end{equation}
depends on the projected semimajor axis, $a \sin i$, where $i$ is the
inclination of the orbital plane, the orbial period, $P_\mathrm{orb}$,
the eccentricity, $e$, the longitude of periastron, $\omega$, and the
true anomaly, $\theta(t)$, which is found by solving Kepler's
equation. Using the orbital parameters\footnote{$\tau =
  \mathrm{MJD}\,54531.7709(0.0603)$, $\omega = 48.67(4)^\circ$, and
  $\dot{\omega} = 0.048(3)^\circ\,\mathrm{yr}^{-1}$ \citep{Li2012a},
  $P_\mathrm{orb} = 24.317037(62)\,\mathrm{d}$ \citep{Bildsten1997a},
  $e = 0.3402(4)$ and $a \sin i = 140.14(16)\,\mathrm{lt-s}$
  \citep{Rappaport1978a}.} of \fu, we calculated the Doppler shift
$\Delta E = E_\mathrm{rest} v(t)/c$, where $E_\mathrm{rest} =
6.4$\,keV is the emission line energy at rest and $c$ the speed of
light, at the time of the \suzaku observations of $\Delta E \approx
3$\,eV. This is orders of magnitude weaker than the observed shift of
210(70)\,eV (epoch~I) or 160(60)\,eV (epoch~II) and, thus, in
contradiction to line emission from an accretion disk.

\subsection{\texttt{CYCLABS} vs.\ \texttt{GABS}}

We have analyzed the spectra using two different models,
\texttt{CYCLABS} (Eq.~\ref{eq:crsf-cyclabs}) and \texttt{GABS}
(Eq.~\ref{eq:crsf-gabs}), for the shape of the CRSFs. The
corresponding best-fit parameters for the three CRSFs detected in the
data are in good agreement among both models (see
Table~\ref{tab:fitpars}): the optical depths and line strengths follow
Eq.~\ref{eq:cyclabs-gabs} within $2\sigma$ confidence and the centroid
energies for the higher harmonics agree within $1.2\sigma$ confidence.
The fundamental CRSF energy in epoch~II seems to differ by $3\sigma$
between \texttt{GABS} and \texttt{CYCLABS}. This is due to the
degeneracy between the centroid energy of the 10\,keV feature and the
fundamental CRSF energy: The MCMC contours clearly show, however, that
the energies agree within the 68\% contour when projected onto the
CRSF energy (Fig.~\ref{fig:MCMCmap}).

Since the CRSF parameters are in agreement between both models and the
corresponding fit goodness do not allow us to favor one of these
models, we use the results of the \texttt{CYCLABS} model in the
following discussions, not least in order to be comparable with
earlier works which used the same model (e.g.,
\citealt{Nakajima2006a}, \citetalias{Mueller2013a}).

The continuum and iron line parameters are in excellent agreement
($\le 1.3\sigma$) when using the \texttt{CYCLABS} or the \texttt{GABS}
model for the shape of the CRSFs. This is expected since the model
narrow features compared to the broad-band X-ray continuum of \fu. We
note a marginal shift of ${\sim}0.15$\,keV in the centroid energy of
the 8.5\,keV Gaussian between both models (see
Fig.~\ref{fig:MCMCmap}), which is due to the detected degeneracy with
the fundamental CRSF.

\subsection{The 10\,keV feature}
\label{sec:discuss:10keV}

Additional spectral components below and around 10\,keV on top of the
underlying cutoff power-law continuum are commonly detected in the
X-ray spectra of highly magnetized neutron star X-ray binaries. In
many cases, these features can be modeled by a black-body component
with temperatures $kT$ between 1 and 2\,keV (see, e.g.,
\citealt{Ballhausen2016a} for KS~1947+300,
\citealt{Caballero2007a,Caballero2013a} for A~0535+26,
\citealt{Kuehnel2013a,Kuehnel2017a} for GRO~J1008$-$57, or
\citealt{Rothschild2017a} for GX~304$-$1). The black-body flux
relative to the power-law seems to depend on luminosity and ranges
from 10\% up to 100\%. In other cases, it is necessary to add a
Gaussian feature in emission or in absorption (see, e.g.,
\citealt{Ferrigno2009a} for \fu, \citealt{Suchy2008a} for Cen~X-3,
\citealt{Vasco2013} for Her~X-1, \citealt{Fuerst2014} for Vela X-1,
and a systematic study in \citealt{Coburn2002a}). These Gaussians,
however, correct some significant, but energetically marginal feature
due to their lower relative fluxes compared to source with black-body
components.

The flux of the Gaussian centered at 8.5\,keV, which we needed to
describe the \suzaku and \rxte spectra of \fu (see
Sect.~\ref{sec:models}), was ${\sim}35$\% and ${\sim}51$\% relative to
the power-law during Epochs~I and II, respectively (see
Table~\ref{tab:fitpars}). This is much too strong to be due to
``photon spawning'' caused by photons produced by electrons excited
into higher Landau levels and de-exciting down into the fundamental
level (see, e.g., \citealt{schwarm:17b} and \citealt{isenberg:98b} for
detailed calculations of the line shape) and comparable to the
black-body flux seen in other sources. Thus, we tried to replace the
Gaussian by such a black-body spectrum. This attempt, however, failed
with an insufficient goodness of the fit ($\chi^2_\mathrm{red} > 2$)
due to strong residuals below 10\,keV. In order to understand this
failed fit Fig.~\ref{fig:bbodyGauss} compares the 8.5\,keV-Gaussian to
a black body with $kT = 2.7$\,keV and the same relative flux as in the
Gaussian of epoch~I. One can see that the width of the Gaussian is
narrower than that of the black body. Since both features contain the
same flux, i.e., their areas are equal, the Gaussian ``sticks out'' on
top of the continuum in contrast to the black body.

We conclude that the Gaussian centered around 8.5\,keV in the spectra
of \fu is peculiar compared to other accreting pulsars: it has a
strong contribution to the overall continuum compared to known
``10\,keV features'' and is incompatible to the black-body components
detected in other sources. This might thus point to a distinct
physical origin. In the remainder of this section, we propose
cyclotron cooling as the physical origin of this Gaussian feature
following discussions and results by \citet{Ferrigno2009a} and
\citet{Farinelli2016a}. In contrast to cyclotron absorption lines,
where photons excite the electrons and are, thus, observed as
absorption features, cyclotron cooling is based on collisional
excitation within the plasma, which results in an additional emission
of photons at the cyclotron energy, i.e., cyclotron emission.

According to \citet{Arons1987a} cyclotron emission can be the dominant
cooling channel in the accretion column once the plasma temperature
and cyclotron energy are comparable. This enables the excitation of
the first Landau level of electrons due to collisions with protons
inside the plasma. The de-excitation will emit a photon at the
cyclotron energy which will effectively cool the plasma. This
cyclotron emission provides seed photons for bulk and thermal
Comptonization inside the accretion column. The self-consistent
physical model by \citet[and references therein]{Becker2007a} uses
analytical approximations in order to calculate the emerging X-ray
spectrum from these seed photons and those from Bremsstrahlung and
black-body emission.

The self-consistent model proposed by \citet[and references
therein]{Becker2007a} was first implemented by \citet{Ferrigno2009a}
and applied to the \textsl{BeppoSAX} spectrum of \fu during its 1999
outburst. It turned out, however, that the assumptions and
restrictions of the \citeauthor{Becker2007a} model are not sufficient to
explain the complex X-ray continuum spectrum of \fu. An extension of
the \citeauthor{Becker2007a} model was developed by
\citet{Farinelli2016a,Farinelli2012a}, who have used some analytical
approximations of the original model in combination with numerical
methods, which allowed them to include variations of the magnetic
field along the column and to approximate the cyclotron emission by a
Gaussian line instead of a Dirac function. These authors applied their
so-called \texttt{COMPMAG} model to the same \textsl{BeppoSAX} data of
\fu as analyzed by \citet{Ferrigno2009a} previously, among
others. They found that their model was able to describe the
0.1--100\,keV spectrum of a few sources, among \fu at different flux
states, \emph{without the introduction of any further emission
  components}\footnote{However, a partial covering medium was needed
  to level out residuals in \fu.}. In particular, a prominent bump in
the spectrum around 9\,keV was due to significant cyclotron emission
predicted by the model. Their best-fit width of cyclotron emission of
2.4\,keV is explained by broadening of the intrinsic cyclotron
emission by, e.g., magnetic field gradients inside the column. Due to
the similarity of the findings by \citet{Farinelli2016a} and the
parameters of our Gaussian component around 8.5\,keV (the ``10\,keV
feature''), we argue that this component accounts for cyclotron emission
as part of our phenomenological model.

Two issues with this scenario remain, which are the difference between
the fundamental cyclotron absorption line energy of $E_\mathrm{CRSF,0}
= 10.2$\,keV and the centroid energy of the cyclotron emission at
$E_\mathrm{10\,keV}{\sim}8.5$\,keV, which is not expected at first
glance, and why cyclotron emission seems to be dominant in \fu but not
in other sources. One explanation of the former issue was put forward by
\citet{Tsygankov2018}, who argue that cyclotron emission cannot escape 
at the resonant energy, owing to the the high scattering cross section. 
Emission is then preferentially detectable at lower energy, at which it can 
more easily escape.
\citet{Ferrigno2009a} proposed that cyclotron
emission appears at higher altitudes in the column than the formation
of the CRSFs, where the magnetic field strength and, thus, the
cyclotron energy is lower than closer to the surface. Assuming a
dipole magnetic field the magnetic field strength, $B(h)$, as a
function of the height, $h$, above the surface is
\begin{equation}\label{eq:dipolefield}
  B(h) = B_0 R^3 / (R+h)^3
\end{equation}
where $B_0$ is the surface magnetic field strength and $R$ the neutron
star's radius. Cyclotron emission is important as long as the
$B$-field energy is comparable to the plasma temperature, which
increases the occurrence of collisions inside the column
\citep{Arons1987a}. The temperature of plasma in the accretion column
is of the order of several keV \citep{Basko1976a,Ferrigno2009a} with a
slow decrease with height (e.g., Fig.~5 of \citealt{Basko1976a} or
\citealt{West2017}). As the rate of production of cyclotron photons
scales as $\exp(-E_\mathrm{cyc}/kT)$ (see eq.~(114) of
\citealt{Becker2007a}), it is plausible that cyclotron emission is
more effective a few hundreds meters above the surface, if one assumes
that cyclotron absorption features are imprinted at the column's base.

We note that conditions for cyclotron emission change as soon as the
magnetic field strength of the neutron star is higher than that of
\fu. A $B$-field more than twice as strong (i.e., a CRSF at
${\gtrsim}21$\,keV), which is the case for most cyclotron line sources
\citep[see, e.g., the review by][]{Staubert2019a}, results in a
cyclotron energy significantly higher than the plasma temperature
throughout the column and, thus, hampering collisional excitation of
the Landau level. If true, this would explain why the spectrum of \fu
is dominated by cyclotron emission in contrast to that of others
sources. We note that the case of GX~304$-$1 at low luminosity might
be different, as discussed in \citet{Tsygankov2018}, because a high
temperature can be reached in a tenuous accretion stream, where
thermal cooling is inefficient and the Landau level can be
collisionally populated.

While cyclotron emission \emph{could} explain both the issues raised
above, we stress that detailed and self-consistent calculations are
needed to fully investigate this idea.

\subsection{Energy dependence of the fundamental CRSF}
\label{sec:discuss:E0}

\begin{figure}
 \includegraphics[width=\columnwidth]{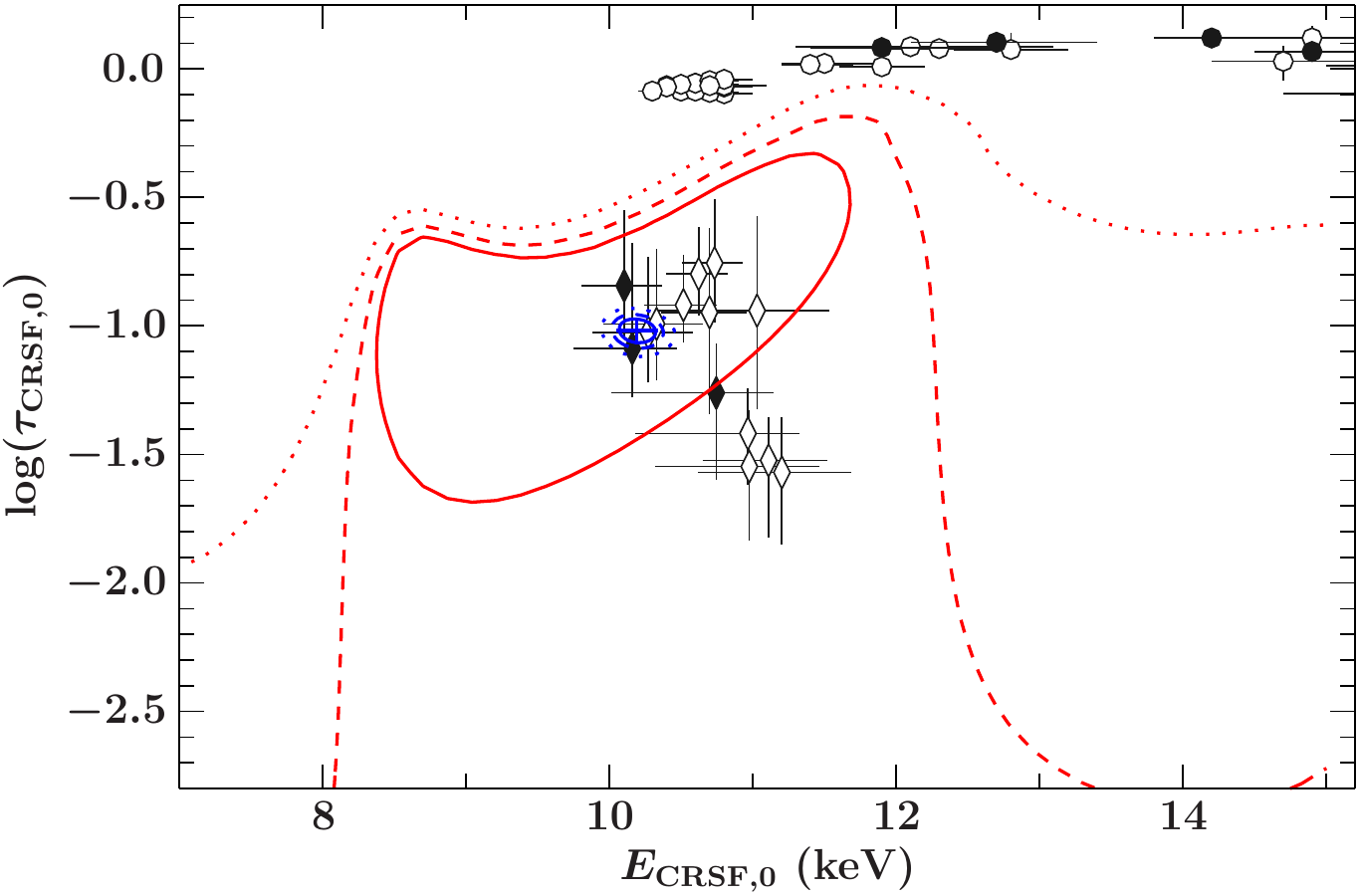}
 \caption{$\chi^2$-contour-map (solid line: 68\% confidence; dashed:
   90\%; dotted: 99\%) between the fundamental CRSF energy,
   $E_\mathrm{CRSF,0}$, and its depth, $\tau_\mathrm{CRSF,0}$, in
   epoch~I (red) and epoch~II (blue). The blue cross marks the
   best-fit values from epoch~II with their respective uncertainties
   (see Table~\ref{tab:fitpars}). The CRSF parameters found by
   \citet[Table~4; circles]{Nakajima2006a} and by \citet[Table~2;
     diamonds]{Mueller2013a} are added for comparison. The filled
   symbols mark parameters obtained at a source luminosity similar to
   Epoch~I and Epoch~II.}
 \label{fig:CRSFmap}
\end{figure}

Previous work on \fu \citep[e.g.,][]{Nakajima2006a, Mueller2010a,
  Li2012a} found an anticorrelation between the fundamental cyclotron
line energy and the X-ray flux. However, \citetalias{Mueller2013a}
found that this apparent anticorrelation depends on the choice of the
continuum model. They argued for an absorbed cutoff power-law with a
10\,keV emission feature, as used in this analysis, for which the
anticorrelation vanishes. As noted in the previous section, the data
gap of \suzaku between XIS and PIN does not allow us to directly study
the behavior of the fundamental cyclotron line between the data
epochs~I and II. In the following, we can nonetheless determine
whether our results are consistent with the anticorrelation between
the fundamental line energy and the source's flux as found by, e.g.,
\citet{Nakajima2006a} or instead, consistent with the uncorrelated
parameters found by \citetalias{Mueller2013a}.

In epoch~I, where we only have \suzaku data available, we had fixed
the energy and depth of the fundamental cyclotron line,
$E_\mathrm{CRSF,0}$ and $\tau_\mathrm{CRSF,0}$, respectively, to the
results of our combined analysis of the \suzaku and \rxte data of
epoch~II (see Sect.~\ref{sec:specana}). In order to investigate an
evolution of the CRSF from epoch~I to epoch~II despite the data gap
between XIS and PIN, we calculated $\chi^2$-contours between these
parameters in epoch~I. The resulting contour-map is shown in
Fig.~\ref{fig:CRSFmap}. The best-fit values of the CRSF energy and
depth of epoch~II (blue cross) are consistent within the $1\sigma$
2D-contour calculated for epoch~I and the possible CRSF energy range
is 8.2--12.2\,keV (90\% confidence). This justifies our approach of
fixing the parameters of the fundamental CRSF parameters in epoch~I to
the best-fit values from epoch~II.

The contour-maps alone do not allow us, however, to favor one of the
two claimed evolutions of the CRSF parameters with the source's flux.
Thus, we compare the parameter combinations enclosed by the contours
with the energies and depth of the fundamental CRSF as found by
\citet{Nakajima2006a} and \citetalias{Mueller2013a} (circles and
diamonds in Fig.~\ref{fig:CRSFmap}, respectively). In particular, we
focus on their parameters derived from observations of \fu at a
similar 3--50\,keV luminosity level (filled symbols in
Fig.~\ref{fig:CRSFmap}) as during epoch~I and epoch~II. Neither the
$\chi^2$-contours of epoch~I nor those of epoch~II are consistent with
the values expected from the anticorrelation as found by
\citet{Nakajima2006a} with at least 99\% confidence. Instead, our
results are in excellent agreement with the findings of
\citetalias{Mueller2013a}. Thus, using the continuum model of
\citetalias{Mueller2013a} to describe the \suzaku-data results in CRSF
energies which are in agreement with expectations from a constant
behavior rather than implying an anticorrelation of the CRSF energy
with luminosity. We note, however, that we cannot exclude a change in
the fundamental CRSF energy between epochs~I and II on the level of a
few keV as a consequence of the data gap between \suzaku-XIS and -PIN
(see the contours for epoch~I in Fig.~\ref{fig:CRSFmap}). To our
knowledge, however, no one has claimed a CRSF-to-luminosity
correlation on this energy scale yet. Although the theory by
\citet{Becker2012a} explains changes of the CRSF energy with the
mass-accretion rate as seen in other accreting pulsars, the
fundamental CRSF energy of \fu around 10--11\,keV is so low that the
regimes of Coulomb and radiation braking after their theory cannot be
distinguished anymore \citep[see Fig.~2 (right) of][]{Becker2012a}.

\subsection{The question of a second fundamental CRSF}

\citet[][I15 for the remainder of this section]{Iyer2015a} analyzed
data from various satellites and instruments during the 2011 outburst
of \fu. In particular they analyzed the simultaneous observations by
\rxte (96032-01-04-00) and \suzaku (406049010), restricted to the time
interval where both observations overlapped (about ${\sim}13$\,ksec).
We analyzed these data as well (epoch~II in this work), but using the
full available on-source time and a second \rxte observation
(96032-01-04-01). Since our best-fit using a common set of parameters
is acceptable (see Table~\ref{tab:fitpars}), we find that restricting
the data to the overlapping interval is not necessary, allowing us to
increase the SNR.

\begin{figure}
 \includegraphics[width=\columnwidth]{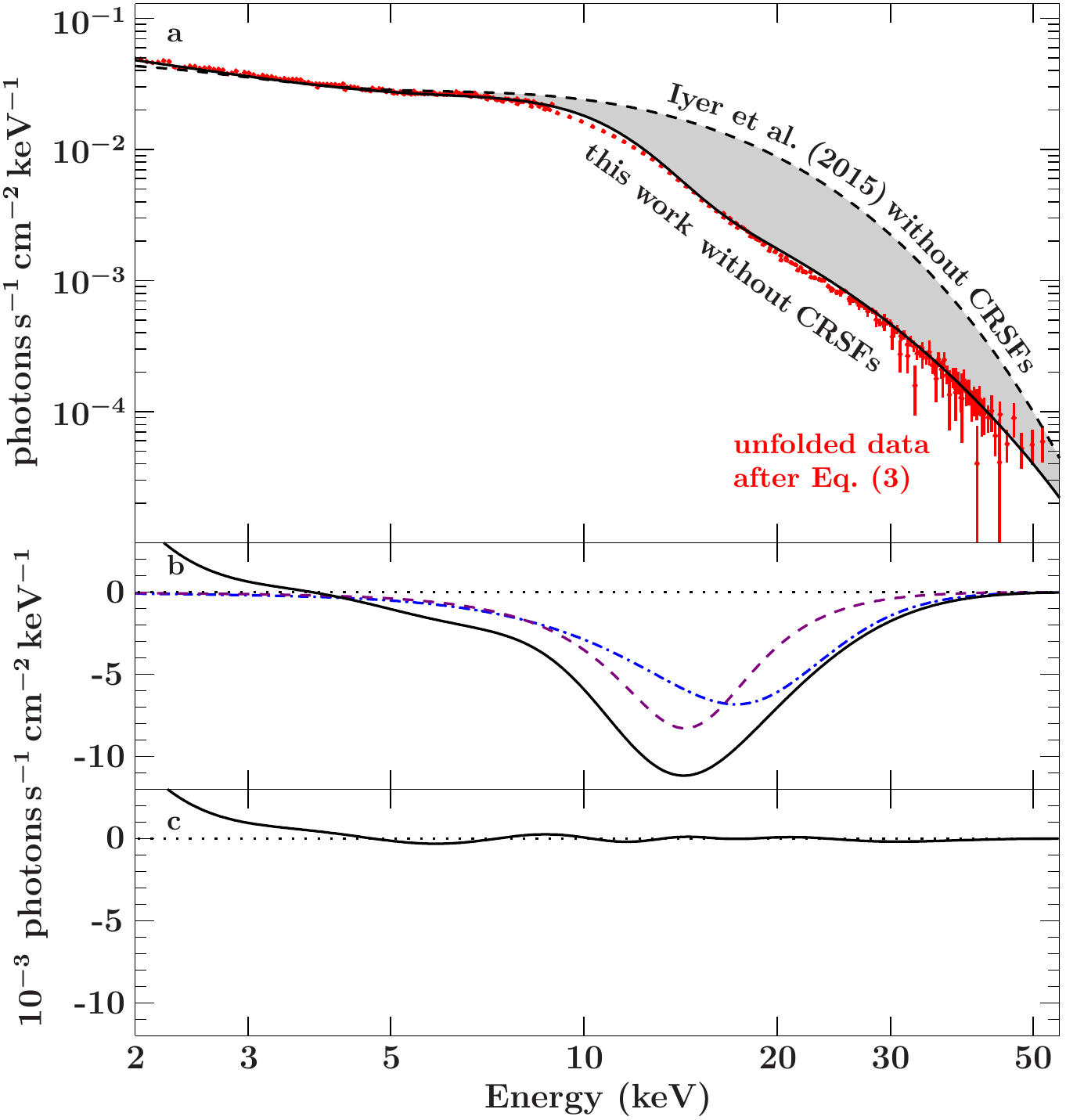}
 \caption{Comparison between the continuum models by \citet{Iyer2015a}
   and this work for epoch~II (\rxte and \suzaku). \textbf{a}
   Unabsorbed continuum model, i.e., without emission lines and CRSFs
   assumed by \citetalias{Iyer2015a} (their best-fit parameters;
   dashed line) and this work (solid line). The continua differ
   significantly between 5 and 50\,keV (gray shaded region). The
   source's flux observed during epoch~II with all instruments,
   unfolded using Eq.~\ref{eq:unfold} and corrected for X-ray
   absorption, is shown for comparison (red data). \textbf{b}
   Difference between the continuum in this work and in
   \citetalias{Iyer2015a} (black solid line). The shapes of the CRSFs
   at 15\,keV (purple dashed line) and the first harmonic around
   20\,keV (blue dashed-and-dotted line) inferred from the parameters
   by \citetalias{Iyer2015a} are shown for comparison. \textbf{c}
   Difference between the continuum-only of this work and the
   \citetalias{Iyer2015a} continuum with their CRSFs at 15\,keV and
   20\,keV included.}
 \label{fig:compareiyer}
\end{figure}

The continuum model chosen by \citetalias{Iyer2015a} differs from our
choice: they used a combination of a low-temperature black body and a
power-law with a high-energy cutoff (\texttt{cutoffpl}, see
Eq.~\ref{eq:cutoffpl}) without the need for a broad Gaussian at
8.5\,keV. Apart from the known Fe K$\alpha$ line and the known CRSFs,
they found absorption-like residuals around $\sim$15\,keV, which they
interpreted as a \emph{second fundamental} cyclotron line. They
attribute this feature to the second magnetic pole of the neutron
star, which might have a higher magnetic field strength than the other
pole due to asymmetries in the magnetic field configuration. Thus,
Iyer et al.\ argued that the superposition of both fundamental lines
might have caused the observed anticorrelation of the 11\,keV line in
the past. However, as described in the previous section, a CRSF at
energies higher than 11\,keV would worsen the best-fit of epoch~II
significantly, i.e., we find no indication for a second fundamental
CRSF at 15\,keV. In order to understand this difference, we compare
and investigate the continuum and CRSF parameters of our work and the
work by \citetalias{Iyer2015a} in the following.

We implemented the model used by \citetalias{Iyer2015a} as defined in
their Section~3 and the parameters given in their Table~4 and 5 for
the simultaneous \suzaku and \rxte observation. Since the detector
response files might have been updated since 2015, we re-fitted their
model to our extracted spectra. The resulting parameters are very
similar to those reported in \citetalias{Iyer2015a}. For the power-law
normalization \citepalias[which is missing in][]{Iyer2015a}, we find a
flux of
$7.9^{+1.0}_{-0.9}\times
10^{-3}\,\mathrm{photons}\,\mathrm{keV}^{-1}\,\mathrm{cm}^{-2}\,\mathrm{s}^{-1}$
at 1\,keV. In contrast to the fit goodness of 688/469 ($\chi^2$/dof)
reported by \citetalias{Iyer2015a} (their Table~4), their model fitted
to our extraction results in a worse goodness of 975/571 (with 1\%
systematic uncertainties added to the full PCA energy range as in
\citetalias{Iyer2015a}). This is due to the higher SNR of our data,
since we did not restrict the observations to the overlapping time
interval. We base the following discussion on their best-fit
parameters \citep[][Table~4 and 5]{Iyer2015a}.

The integrated flux over the continuum model of \citetalias{Iyer2015a}
while excluding the CRSF components is about twice as high as over our
model ($8.6{\times}10^{-9}\,\ergs$ vs.\ $4.1{\times}10^{-9}\,\ergs$ in
the 1--100\,keV energy range). This is due to a significantly higher
continuum flux between 5 and 50\,keV (see gray shaded region in
Fig.~\ref{fig:compareiyer}a) caused by the negative photon
index\footnote{The photon index, $\Gamma$, as defined in
  Eq.~\ref{eq:cutoffpl} and implemented in XSPEC is found to be
  positive for almost all accreting pulsars \citep[see,
  e.g.,][]{Bildsten1997a}.} of $\Gamma = -1.25$ found by these
authors, which results in an \emph{increasing} photon flux with
energy, ignoring the exponential cut-off. For illustration purposes we
included in Fig.~\ref{fig:compareiyer}a the model-independent photon
flux observed during epoch~II, $\bar S(h)$. Following
\citet{nowak:05a} and \citet[][Sect.~7.2]{ISISmanual}, the unfolded
flux in spectral bin $h$ is given by
\begin{equation}\label{eq:unfold}
  {\bar S(h)} \equiv \left( C(h) - B(h) \right) \,\times\, \left( t~\int_{\Delta E(h)}\D
    E~R(h,E)\,A(E) \right)^{-1}
\end{equation}
where $C(h)$ is the observed counts in this bin, $B(h)$ is the
background count, $t$ is the exposure time, $\Delta E(h)$ is the
energy range contributing to detector bin $h$, $A(E)$ is the detector
effective area at energy $E$ and $R(h,E)$ is the redistribution
function of the detector. Eq.~\eqref{eq:unfold} assumes that the
source flux is constant in the interval $\Delta E(h)$, which is a
reasonable approximation for photon energies $E \gtrsim 1$\,keV and
typical CCD detector resolution \citep[see Sect.~7.2
of][]{ISISmanual}. As shown in Fig.~\ref{fig:compareiyer}a, the
observed flux does not agree with the continuum model of
\citetalias{Iyer2015a}, while it follows our continuum model almost
perfectly with the exception of small deviations as expected at the
CRSF energies (see Fig.~\ref{fig:spec}c for the residuals). Since
\citetalias{Iyer2015a} claim to have found a good description of the
spectra, further components in their full model have to have reduced
their continuum flux to the observed values. As shown in
Fig.~\ref{fig:compareiyer}b, the difference between the continua
appears as a broad line-like absorption feature centered at
${\sim}15$\,keV. We find that the shape of the second fundamental CRSF
at 15\,keV and the first harmonic around 20\,keV both claimed by
\citetalias{Iyer2015a} (see red and blue lines in
Fig.~\ref{fig:compareiyer}b) are very similar in width and depth
compared to the difference of the continua (black line). Once the
continuum model by \citetalias{Iyer2015a} is modified by these two
CRSFs, the difference to our continuum model is almost gone (see
Fig.~\ref{fig:compareiyer}c). In fact, as shown in
Table~\ref{tab:fluxcrsf} these two components absorb ${\sim}65\%$ of
the total flux of the baseline continuum, while in our model all CRSF
components absorb only ${\sim}4\%$ in total. Thus, we conclude that
the excessive flux of the continuum model by \citetalias{Iyer2015a} is
erroneously corrected by their claimed CRSFs. This argument is
consistent with one made by some of us earlier in relation to the
interpretation of the CRSFs of \fu \citep{Mueller2013a}.

\begin{table}
  \centering
  \caption{Comparison of the photon fluxes absorbed by the fundamental
    cyclotron line, E$_0$, and its higher harmonics, E$_1$ and E$_2$,
    during epoch~II as derived from the models by \citet{Iyer2015a}
    and this work.}
  \label{tab:fluxcrsf} 
  \renewcommand{\arraystretch}{1.3}
  \setlength{\tabcolsep}{3pt}
  \begin{tabular}{lllll}
    \hline\hline
    \quad type & E$_0$ & 2$^\text{nd}$~E$_0$ & E$_1$ & E$_2$ \\
    \hline
    \csname @@input\endcsname linefluxes
    \hline
  \end{tabular}
  \tablefoot{
    Uncertainties and upper limits are given at the 90\% confidence level.\\
    \tablefoottext{a}{Absorbed flux by the CRSF in
      $10^{-12}$\,erg\,s$^{-1}$\,cm$^{-2}$.\\}
    \tablefoottext{b}{Absorbed flux by the CRSF relative to the
      continuum flux in the 1--60\,keV energy range.\\}
    \tablefoottext{c}{residual flux $r = \int_{-W_\text{CRSF}}^{+W_\text{CRSF}} {\D}E\,F_\mathrm{\nu}(E - E_\text{CRSF}) / F_0(E - E_\text{CRSF})$ after Eq.~\ref{eq:residualflux}.\\}
  }
\end{table}

\section{Conclusions and Summary}\label{sec:summary}

In this work we analyzed two sets of spectral data for the giant
outburst of \fu in 2011, which were taken during the fading phase of
the outburst. The spectra can be described with an absorbed cutoff
power-law, modified by a Gaussian emission feature around 8.5\,keV,
narrow emission lines from neutral and ionized iron, and three
cyclotron features at $\sim$10\,keV, $\sim$21, and 33\,keV,
respectively.

A main aim of this work is to study the behavior of the CRSFs in
comparison with previous findings for this source
\citep[e.g.,][]{Nakajima2006a,Tsygankov2007a,Mueller2010a,Li2012a}. In
contrast to claims in earlier publications, \citetalias{Mueller2013a}
showed that the energy of the fundamental CRSF does not exhibit a
correlation with the source flux. They argue that the previously
claimed anticorrelation is due to the usage of an insufficient
continuum model, which causes the cyclotron line components to
erroneously correct for the continuum shape. \citet{boldin2013a} have
confirmed that the choice of the continuum model strongly influences
the best-fit parameters of the cyclotron lines. For the \suzaku data
analyzed here, we reject the CRSF energies as expected from an
anticorrelation (see Sect.~\ref{sec:discuss:E0}) and, thus, confirm
the results of \citetalias{Mueller2013a}.

\citet{Iyer2015a} proposed that the discrepancy in the literature
concerning the behavior of the fundamental CRSF could be caused by
flux-dependent contributions of two fundamental cyclotron lines,
located at $\sim$10.5\,keV and $\sim$15\,keV, respectively. These
authors base this conclusion on an analysis of simultaneous \suzaku
and \rxte data for the 2011 outburst. Applying the model by
\citetalias{Mueller2013a} to the same data (epoch~II), we find,
however, no evidence of a second fundamental cyclotron line. Rather,
we have shown that the line-like feature at 15\,keV seen by
\citet{Iyer2015a} is again caused by the choice of the
phenomenological continuum model (see Fig.~\ref{fig:compareiyer}).
Based on the unphysically large fraction of the continuum flux
absorbed by the CRSF features claimed by \citet{Iyer2015a}, we
therefore conclude that there is no second fundamental line in the
spectrum of \fu.

On a more general note, we strongly suggest checking all cyclotron
line parameters found by fitting phenomenological models to observed
X-ray spectra against physically expected parameter ranges. For
example, if one can assume that the magnetic field is constant in the
region in which the CRSF is formed, then sophisticated Monte Carlo
simulations in order to derive accurate cross-sections for cyclotron
scattering \citep{Schwarm2017a} show that the width, $\Delta E$, of a
fundamental CRSF with the centroid energy $E$ can be approximated by
the full Doppler width for thermal cyclotron line broadening
\citep[see also][]{MeszarosNagel1985a},
\begin{equation}
  \frac{\Delta E}{E} = \sqrt{8 \ln 2 \frac{k_\mathrm{B} T}{m_\mathrm{e} c^2}} \cos \theta
\end{equation}
where $T$ is the electron temperature, $\theta$ is the angle between
the incident photon trajectory and the magnetic field, $m_\mathrm{e}$
is the electron rest mass, $c$ the speed of light, and $k_\mathrm{B}$
is the Boltzmann constant. Assuming $k_\mathrm{B} T = 7.94$\,keV,
based on the modeling of the X-ray spectrum of \fu with a bulk and
thermal Comptonization model \citep{Ferrigno2009a}, we find a relative
width of 30\% at maximum ($\cos \theta = 1$), which translates to
${\sim}3$\,keV for the CRSFs in \fu. In case of higher harmonic lines,
the thermally averaged cross-sections calculated by
\citet{Schwarm2017a} do not indicate significantly broader lines
compared to the fundamental line. Cyclotron lines broader than a few
keV can be explained by, e.g., a $B$-field gradient along the line
forming region and specific viewing angles or velocity gradients
\citep[e.g.,][]{nishimura13a,nishimura14a,nishimura19a,poutanen2013a}.
These effects should, however, influence all CRSFs present in the
spectrum. In summary, care must be taken as soon as any CRSF component
is significantly broader than ${\Delta E}/{E} > 30\%\,\text{--}\,40\%$
or when the relative widths of multiple detected CRSFs are
significantly different from each other.

A similar check is possible for the depth of the cyclotron line.
As shown by \citet{harding1991a}, for photons close to the resonant
frequency the first order absorption cross section is a very good
approximation for the inelastic one-photon scattering cross section,
and therefore magnetic Compton scattering in the line core can be
approximated as an absorption process. Therefore, using the fitted
value for $\tau_\mathrm{CRSF}$, i.e., the optical depth in the core of
the cyclotron line, we can estimate the residual flux, $r$, around the
position of a cyclotron line at energy $E_\mathrm{CRSF}$ to
\begin{equation}\label{eq:residualflux}
  r = \int_{-W_\text{CRSF}}^{+W_\text{CRSF}} {\D}E\,
  \frac{F_\mathrm{\nu}(E-E_\text{CRSF})}{F_0(E-E_\text{CRSF})} \sim 
  e^{-\tau_\text{CRSF}}\quad{.}
\end{equation}
where $W_\text{CRSF}$ is the width of the cyclotron line,
$F_\mathrm{\nu}$ is the emergent flux including cyclotron scattering,
and $F_0$ the input continuum flux without a CRSF. In the Monte Carlo
simulations performed by \citet{Schwarm2017a}, only 1--10\% of the
initial photons undergo resonant scattering, i.e., $r \gtrsim
90\%$. Consequently, we would expect $\tau_\mathrm{CRSF} \lesssim 0.1$
for the observed optical depths\footnote{Note that the definition of
  the optical depth depends on the chosen phenomenological absorption
  model \citep[e.g.,][and references therein]{Staubert2019a}. For
  instance, the optical depths of the \texttt{CYCLABS}- and
  \texttt{GABS}-model (see Eqs.~\ref{eq:crsf-cyclabs} and
  \ref{eq:crsf-gabs}) are linked by Eq.~\ref{eq:cyclabs-gabs}.}. From
our best-fit model of the spectra of Epoch~II, the calculated residual
fluxes after Eq.~\ref{eq:residualflux} are $\ge 78\%$ for all three
detected cyclotron lines (see Table~\ref{tab:fluxcrsf}). These values
together with the fitted optical depths of $\tau_\mathrm{CRSF} \le
0.24$ (see Table~\ref{tab:fitpars}) are in excellent agreement with
the expectations from the simulations by
\citet{Schwarm2017a}. Although spawned photons from higher harmonics,
which originate from multi-photon scattering and from the radiative
deexcitation of electrons excited into higher levels, can affect the
shape of the fundamental line \citep{isenberg:98b}, they only can
decrease its depth. We note that residual fluxes down to $r \gtrsim
36\%$ have been observed for GX~304$-$1 \citep{Rothschild2017a}, which
can be obtained by, e.g., a higher optical depth at the line forming
region. However, it is not expected that the CRSFs absorb a
significant fraction of the total broad-band continuum flux, which is
the case for the model by \citet{Iyer2015a} and conclude that the CRSF
parameters found by these authors are due to a degeneracy between the
continuum modeling and the CRSF modeling when unphysical values for
the CRSF are allowed.

In summary, the choice of a phenomenological model for the X-ray
spectrum of an accreting neutron star may cause strong discrepancies
relative to the theoretically expected values for the CRSF parameters.
In particular, an erroneous shape or an overestimation of the X-ray
flux is sometimes corrected by the introduction of further, strong
absorption components, which are probably not real. We stress that we
do not claim that our choice of the phenomenological continuum model
for \fu describes its true X-ray spectrum best, however, in contrast
to other continua the model applied here yields CRSF parameters which
are consistent with theoretical expectations. These conclusions
desperately call out for a self-consistent model for both the spectral
continuum and the cyclotron resonant scattering features.

\begin{acknowledgements}
  We acknowledge funding by the European Space Agency under contract
  number C4000115860/15/NL/IB, by the Bundesministerium f\"ur
  Wirtschaft und Technologie under Deutsches Zentrum f\"ur Luft- und
  Raumfahrt grants 50OR0808, 50OR0905, 50OR1113, and 50OR1207, and by
  the Deutscher Akademischer Austauschdienst. MTW is supported by the
  NASA Astrophysical Data Analysis Program and the Chief of Naval
  Research. VG is supported through the Margarethe von Wrangell
  fellowship by the ESF and the Ministry of Science, Research and the
  Arts Baden-W\"urttemberg. SMN and JMT acknowledge Spanish Ministerio
  de Ciencia, Tecnolog\'ia e Innovaci\'on (MCINN) through the grant
  ESP2016-76683-C3-1-R and ESP2017-85691-P, respectively. We thank
  John E. Davis for the development of the \texttt{SLxfig} module,
  which was used to create all figures in the paper. We thank the
  anonymous referee for her/his constructive comments on how to
  improve our paper.
\end{acknowledgements}

\begin{appendix}

\section{\suzaku-XIS and \rxte-PCA calibration}

\begin{figure}[h!]
  \includegraphics[width=\columnwidth]{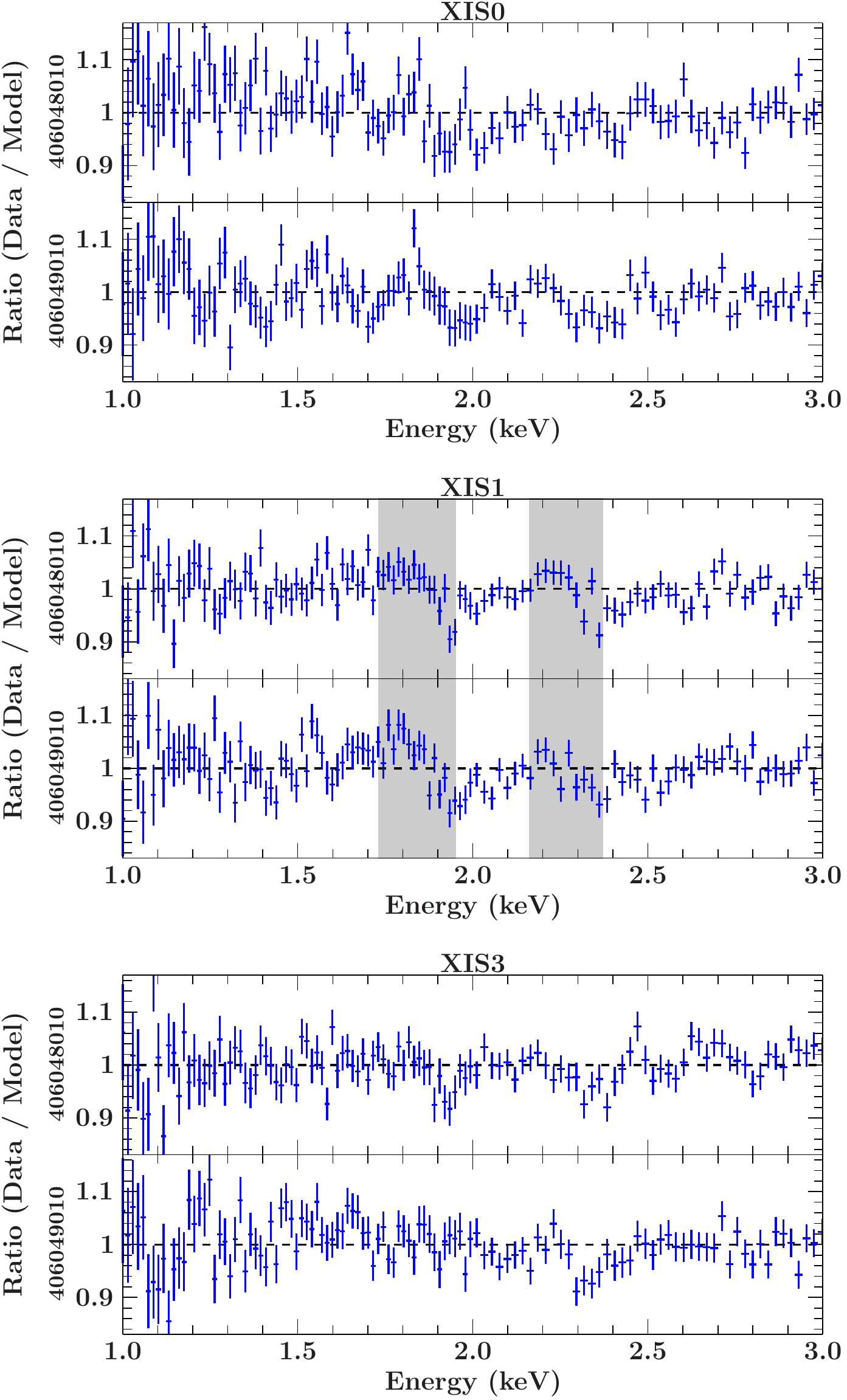}
  \caption{Effect of the updated \suzaku-XIS calibration (20181010) on
    the residuals around the Si- and Au-edges. The residuals were
    calculated from our best-fit fit to epoch~I (upper panels) and
    epoch~II (lower panels) as listed in Table~\ref{tab:fitpars}. The
    energy ranges in gray mark the bins, which were ignored in the
    XIS1-spectra during spectral analysis (Sect.~\ref{sec:specana}).
    For display purposes, the channel binning has been reduced by a
    factor of 2 than as described in Sect.~\ref{sec:obs}.}
  \label{fig:XIScalib}
\end{figure}

\newpage

\begin{figure}[h!]
  \includegraphics[width=\columnwidth]{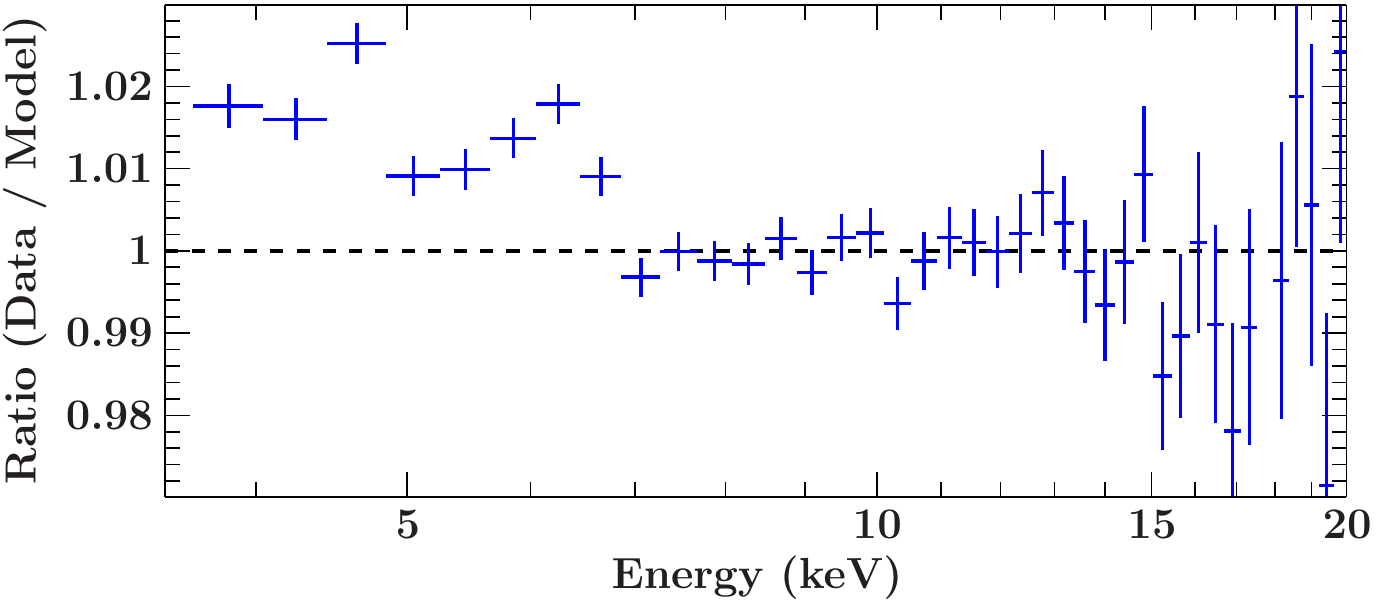}
  \caption{Ratio of the combined \rxte-PCA spectra to our best-fit
    model for epoch~II. Features on a 1--2\% level are visible below
    7\,keV, which existence is not confirmed by the \suzaku-XIS spectra
    (see Fig.~\ref{fig:spec}). Similar features have been seen in
    combined PCA spectra of the Crab pulsar \citep[Fig.~8]{Garcia2014a}
    or of GRO~J1008$-$57 \citep[Fig.~3]{Kuehnel2016a}.}
  \label{fig:PCAcalib}
\end{figure}

\section{The 10\,keV feature vs.\ a black body}

\begin{figure}[h!]
  \includegraphics[width=\columnwidth]{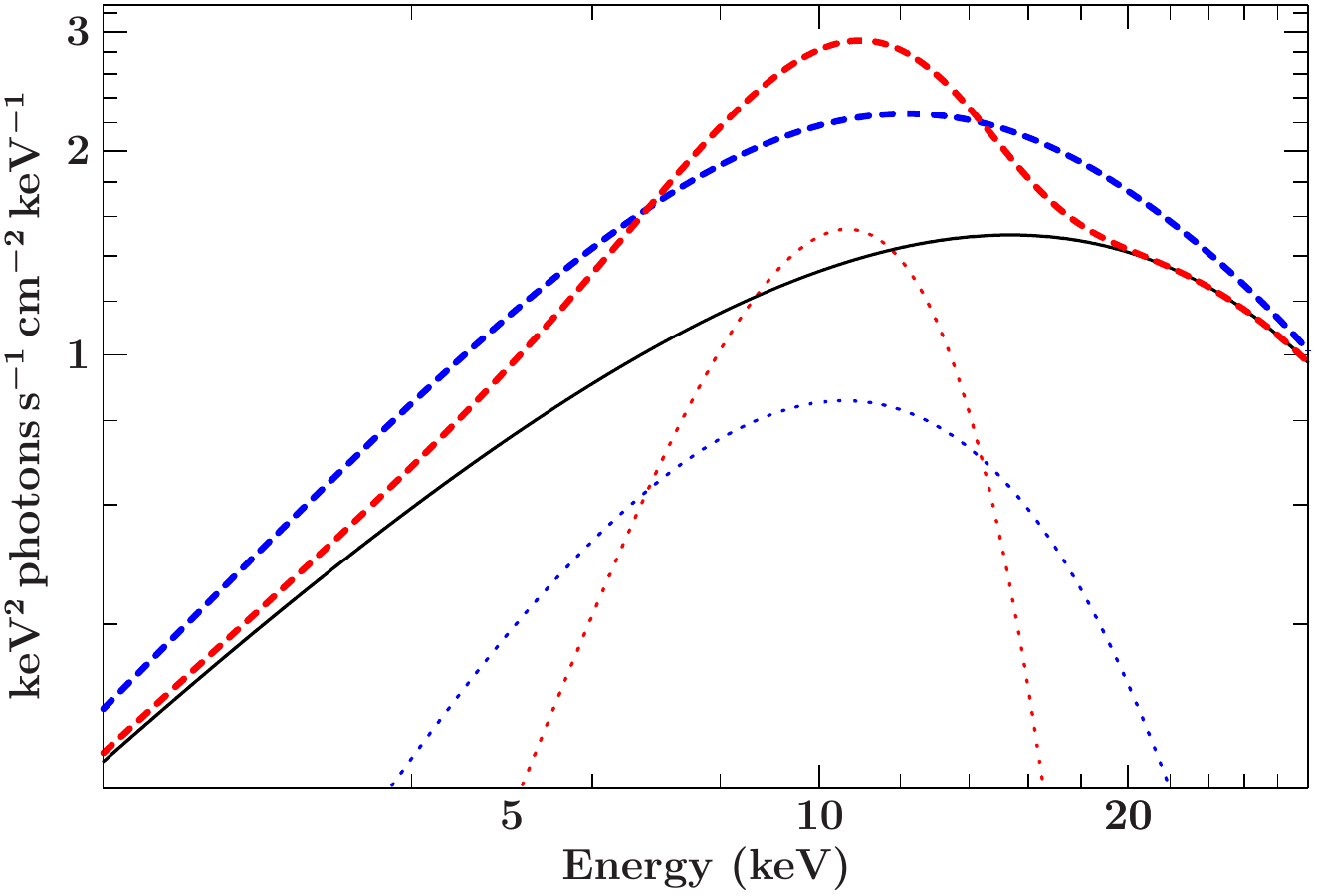}
  \caption{Comparison of the spectral shapes of the 8.5\,keV Gaussian
    (red dotted line) and a black body with $kT = 2.7$\,keV (blue
    dotted line). The total flux is the same for both components. When
    added to the cutoff power-law continuum (black solid line) the
    Gaussian is much more prominent (red dashed line) than the black
    body (blue dashed line). Note that the spectra are shown as
    radiated power ($\nu F_\nu$) over the energy $E = h \nu$.}
  \label{fig:bbodyGauss}
\end{figure}
  
\end{appendix}

\end{document}